\documentclass[twocolumn,showpacs,preprintnumbers,amsmath,amssymb,aps,prl]{revtex4-1}

\usepackage{graphicx}
\usepackage{dcolumn}
\usepackage{bm}
\usepackage{amsmath,amssymb}
\usepackage{epstopdf}
\usepackage{hyperref}
\usepackage{braket}
\usepackage{physics}
\usepackage{footnote}
\usepackage{threeparttable}
\usepackage{tabularx}

\newcommand{\ts}{\textsuperscript}
\newcommand{\vect}[1]{\boldsymbol{\mathbf{#1}}}
\newcommand{\graphsize}{0.45}

\newcolumntype{R}{>{\raggedleft\arraybackslash}X}
\newcolumntype{L}{>{\raggedright\arraybackslash}X}
\newcolumntype{C}{>{\centering\arraybackslash}X}

\begin{document}

\preprint{}

\title{Acoustic Deformation Potentials of $n$-Type PbTe from First Principles}
\author{Aoife R. Murphy\textsuperscript{1,2}}
\author{Felipe Murphy-Armando\textsuperscript{2}}
\author{Stephen Fahy\textsuperscript{1,2}}
\author{Ivana Savi\'c\textsuperscript{2}}
\email{ivana.savic@tyndall.ie}
\affiliation{\textsuperscript{\normalfont{1}}Department of Physics, University College Cork, College Road, Cork, Ireland}
\affiliation{\textsuperscript{\normalfont{2}}Tyndall National Institute, Dyke Parade, Cork, Ireland}

\date{\today}

\begin{abstract}

We calculate the uniaxial and dilatation acoustic deformation potentials, $\Xi^{\text{L}}_{u}$ and $\Xi^{\text{L}}_{d}$, of the conduction band L valleys of PbTe from first principles, using the local density approximation (LDA) and hybrid functional (HSE03) exchange-correlation functionals. We find that the choice of a functional does not substantially affect the effective band masses and deformation potentials as long as a physically correct representation of the conduction band states near the band gap has been obtained. Fitting of the electron-phonon matrix elements obtained in density functional perturbation theory (DFPT) with the LDA excluding spin orbit interaction (SOI) gives $\Xi^{\text{L}}_u = 7.0$~eV and $\Xi^{\text{L}}_d = 0.4$~eV. Computing the relative shifts of the L valleys induced by strain with the HSE03 functional including SOI gives $\Xi^{\text{L}}_u = 5.5$~eV and $\Xi^{\text{L}}_d = 0.8$~eV, in good agreement with the DFPT values. Our calculated values of $\Xi^{\text{L}}_u$ agree fairly well with experiment ($\sim 3-4.5$~eV). The computed values of $\Xi^{\text{L}}_d$ are substantially smaller than those obtained by fitting electronic transport measurements ($\sim 17-22$~eV), indicating that intravalley acoustic phonon scattering in PbTe is much weaker than previously thought.

\end{abstract}


\maketitle

%%%%%%%%%%%%%%%%%%%
% Introduction

\section{I. Introduction}

It has recently become possible to calculate electronic mobility of bulk semiconductors from first principles, without any empirical parameters \cite{Murphy-Armando2008,Restrepo2009,Wang2011,Qiu2015,Zhou2015,Li2015a,Fiorentini2016,Zhou2016a,Gunst2016,Liu2017a,Markussen2017,Ma2018,Ponce2018}. Nevertheless, the accuracy of these first principles methods based on density functional theory (DFT) is still not well established. Electronic mobility of a semiconductor is determined by carrier effective masses and scattering rates of the electronic states near the band gap \cite{Ziman1960,Mermin1976}. However, band gaps are typically underestimated using the standard approximations for the exchange-correlation energy in DFT \cite{Perdew2009}. This may lead to the inaccurate description of the curvatures of the conduction and valence bands near the energy gap, particularly for direct narrow gap semiconductors. Consequently, it is essential to correctly describe the electronic bands near the gap in order to reliably compute electronic mobility of semiconductors. 

The issue of the accurate representation of the electronic states relevant for transport is particularly severe in PbTe, which is one of the most efficient thermoelectric materials in the mid-temperature range ($500-900$ K) \cite{Pei2011,Pei2011a,Lalonde2011,Pei2012}. PbTe has a direct narrow band gap at four equivalent L points \cite{Bauer2003, Dalven1974, Tsang1971}. When spin orbit interaction (SOI) is taken into account  using the local density approximation (LDA) in DFT, the direct band gap at the L point is underestimated to such a degree that the conduction and valence bands invert and mix heavily \cite{Hummer2007, Wei1997, Svane2010a}. This is termed a ``negative band gap'', and results in the charge carrier effective masses in poor agreement with experiment \cite{Hummer2007, Svane2010a}. Such issues in reproducing the electronic band structure of PbTe indicate that it is much more challenging to model its electronic thermoelectric transport properties from first principles \cite{Singh2010,Parker2013,Chen2013,Song2017} than the lattice thermal conductivity \cite{Shiga2012, Skelton2014, Romero2015, Murphy2016, Murphy2017}.

The strength of electron-phonon scattering in PbTe also remains elusive due to the outlined issues in characterizing the electronic band structure. In particular, the strength of intravalley acoustic phonon scattering, which depends on the curvature of the conduction and valence band edges, is not well determined. Intravalley acoustic phonon scattering may be described with deformation potential theory \cite{Bardeen1950,Herring1956}, in which its strength is determined by an effective deformation potential $\Xi$. Empirically, this parameter can be determined by choosing a transport model with selected scattering mechanisms and fitting to electronic transport measurements. In the case of $n$-type PbTe, this approach yielded $\Xi^{\text{L}}=22\pm 2$ eV for the conduction band minima at L~\cite{Lalonde2011, Ravich1971, Ravich1971a}. However, for the L valleys in the cubic structure, there are two linearly independent deformation potentials: uniaxial and dilatation deformation potentials, $\Xi^{\text{L}}_{u}$ and $\Xi^{\text{L}}_{d}$ \cite{Herring1956}. The values of $\Xi^{\text{L}}_u$ can be obtained directly from piezoresistance or ultrasonic measurements, yielding $\Xi^{\text{L}}_u\sim 3-4.5$~eV~\cite{Ravich1971a, Ravich1971}. Combining these values of $\Xi^{\text{L}}$ and $\Xi^{\text{L}}_{u}$ gives $\Xi^{\text{L}}_d\sim 17-22$~eV \cite{Ravich1971, Ravich1971a}. In contrast, early empirical pseudopotential calculations found $\Xi_u^{\text{L}}=8.3$ eV and $\Xi_d^{\text{L}}=-4.4$ eV \cite{Ferreira1965}. A recent first principles study reported only the electronic scattering rates of PbTe resolved by each phonon mode, which were calculated using the electronic band structure with the negative direct gap at L~\cite{Song2017}. 

In this paper, we determine the acoustic deformation potentials of the conduction band L valleys of PbTe using first principles and state-of-the-art hybrid exchange-correlation functionals.~We verify that the LDA excluding SOI and the hybrid HSE03 functional including SOI capture a positive band gap and the effective electron masses in excellent agreement with experiment, in contrast to the LDA including SOI. We show that different exchange-correlation functionals give similar band masses and deformation potentials, as long as they correctly describe the character of the conduction band states near the band edges. We calculate $\Xi_{u}^{\text{L}}=7.0$ eV and $\Xi_{d}^{\text{L}}=0.4$ eV by fitting the electron-phonon matrix elements obtained using density functional perturbation theory (DFPT) with the LDA excluding SOI. We also calculate deformation potentials from relative band shifts due to strain following the approach of Van de Walle and Martin \cite{VandeWalle1986, VandeWalle1989, VandeWalle1987, VandeWalle1989a}, which allows us to compute deformation potentials using hybrid functionals. To the best of our knowledge, DFPT implementations that link with hybrid functionals are still unavailable. Using the approach of Van de Walle and Martin, we obtain $\Xi_{u}^{\text{L}}=5.5$ eV and $\Xi_{d}^{\text{L}}=0.8$~eV with the HSE03 functional including SOI, which are close to our results using the same approach and the LDA without SOI ($\Xi_{u}^{\text{L}}=8.0$ eV and $\Xi_{d}^{\text{L}}=0.5$~eV), as well as our DFPT results. Contrary to the values determined  empirically from transport measurements \cite{Lalonde2011,Ravich1971, Ravich1971a}, we find that $\Xi_{d}^{\text{L}}\sim 0.5$ eV, and the effective deformation potential is approximately proportional to $\Xi^{\text{L}}\sim 6$ eV. This finding is of considerable significance in understanding the material properties that determine carrier transport in PbTe.

%%%%%%%%%%%%%%%%%%%
%  Methodology

\section{II. Methodology}

Within the low temperature and low carrier concentration regime, the four conduction band minima at the L points in PbTe can be well described by a parabolic energy dispersion with respect to the wavevector $\vect{k}$:
\begin{equation}
E(\vect{k}) = E_c + \frac{\hbar^2 \Delta k^2_{\parallel}}{2 m_{\parallel}^{*}} + \frac{\hbar^2 \Delta k^2_{\perp}}{2 m_{\perp}^{*}}.
\label{ch5_parabands}
\end{equation}

\noindent $E_c$ is the energy of the band edge, $\hbar$ is the Planck constant, and $\Delta k_{\parallel}$ and $\Delta k_{\perp}$ are the parallel and perpendicular components of the wavevector $\vect{k}-\vect{k}_{\text L}$, where $\vect{k}_{\text L}$ corresponds to the L valley minimum.  Charge carriers behave as free electrons with effective masses $m_{\parallel}^{*}$ or $m_{\perp}^{*}$ depending on the direction of motion: along the longitudinal axis parallel to $\vect{k}_{\text L}$ or along the transverse axes perpendicular to $\vect{k}_{\text L}$, respectively. We verify this assumption and extract the values of $m_{\parallel}^{*}$ and $m_{\perp}^{*}$ from first principles calculations of the electronic band structure, along with values of the band gap.

To calculate deformation potentials due to long-wavelength acoustic phonons \cite{Bardeen1950,Herring1956}, we follow the procedure outlined by Fischetti and Laux \cite{Fischetti} and Murphy-Armando and Fahy \cite{Murphy-Armando2008}. We may describe the interaction between electrons and acoustic phonons with a slowly varying potential dependent on the deformation potential tensor $\Xi_{\alpha\beta}$ \cite{Ridley1999, Herring1956} as:
\begin{equation}
H_{ep} = \sum_{\alpha\beta} \Xi_{\alpha\beta}\varepsilon_{\alpha\beta}(\vect{r}),
\label{ch5_hamiltonian_acoscatt}
\end{equation}

\noindent where $\alpha$ and $\beta$ are Cartesian coordinates, and $\varepsilon(\vect{r})$ is the local strain tensor at $\vect{r}$ \cite{Yu2005}. Considering the L and X valleys of a cubic material, the deformation potential tensor consists of two linearly independent terms, $\Xi_d^{\nu}$ and $\Xi_u^{\nu}$, $\nu\in$\{L, X\} \cite{Herring1956}. $\Xi_d^{\nu}$ is the dilatation deformation potential, and represents the shift in band energy due to a dilatation in the direction normal to the axis of the valley $\nu$ \cite{Herring1956}. $\Xi_u^{\nu}$ is the uniaxial deformation potential, and corresponds to the band shift due to a uniaxial stretch along the direction of the valley $\nu$, and a corresponding compression perpendicular to this direction so that volume is preserved \cite{Herring1956}. In this case, the electron-phonon coupling Hamiltonian can be written as \cite{Herring1956, Yu2005}:
\begin{equation}
H_{ep} = \Xi_{d}^{\nu} Tr[\vect{\varepsilon}(\vect{r})] + \Xi_{u}^{\nu}(\hat{\vect{k}}_{\nu} \cdot \vect{\varepsilon}(\vect{r}) \cdot \hat{\vect{k}}_{\nu}).
\label{ch5_elphHamiltonian_final}
\end{equation}

\noindent where $\hat{\vect{k}}_{\nu}$ is a unit vector parallel to the $\vect{k}$-vector of the valley $\nu$.

We calculate the uniaxial and dilatation deformation potentials for the conduction band L valleys of PbTe from first principles by linking the electron-phonon coupling Hamiltonian from the deformation potential theory with that of DFPT. Within DFPT, the electron-phonon matrix element for an electron scattering event from a state $\vect{k}$ and band $n$ to a state $\vect{k}{'}$ and band $m$ via a phonon with wavevector $\vect{q}$ and branch index $s$ can be defined as \cite{Giustino2017}:
\begin{equation}
\begin{aligned}
H_{mn}(\vect{k}; \vect{q}s) = \sum_{ b }\sum_{\alpha} & \left( \frac{m_{c}}{m_{b}} \right)^{\frac{1}{2}} e^{s}_{\alpha  b }(\vect{q}) \\ & \times  \bra{u_{m\vect{k}+\vect{q}}} \partial_{\alpha b ,\vect{q}}v^{\text{KS}}  \ket{u_{n\vect{k}}}_{\text{uc}},
\label{ch2_elphmatelement}
\end{aligned}
\end{equation}
\noindent where  $e^{s}_{\alpha b}(\vect{q})$ is the $\alpha$\ts{th} Cartesian component of the phonon eigenvector for an atom $b$ with mass $m_b$. $m_{c}$ is an arbitrary reference mass \cite{Giustino2017}, and we choose it to be equal to the mass of the unit cell for consistency with deformation potential definitions \cite{Vandenberghe2015}. The subscript ``uc'' in Eq.~\eqref{ch2_elphmatelement} indicates that the integral is carried out within one unit cell. $u_{n\vect{k}}$ is normalized to unity in the unit cell, and is the lattice periodic part of the wavefunction $\psi_{n\vect{k}}$ expressed in Bloch form as $N_l^{-1/2}u_{n\vect{k}}e^{i\vect{k}\cdot\vect{r}}$, where $N_l$ is the number of primitive cells. $\partial_{\alpha b ,\vect{q}}v^{\text{KS}}$ is the lattice periodic part of the perturbed Kohn-Sham potential expanded to first order in the atomic displacement, see Ref.~\cite{Giustino2017} for further details.~We note that the above definition of the electron-phonon matrix element is not the typical one given in the DFPT literature, which corresponds to Eq.~\eqref{ch2_elphmatelement} multiplied by  the ``zero-point'' displacement amplitude,  $l_{\vect{q}s} = (\hbar/2m_{c}\omega_{\vect{q}s})^{1/2}$ \cite{Giustino2017,Restrepo2009, Fiorentini2016}. 

To calculate deformation potentials using DFPT, we fit the electron-phonon matrix elements given by Eq.~\eqref{ch2_elphmatelement} in the limit of $\vect{q}\rightarrow 0$ with the Fourier transform of the deformation potential Hamiltonian of Eq.~\eqref{ch5_elphHamiltonian_final}:
\begin{equation}
H_{ep}  =  \Xi^\nu_d {\bf e}^s({\bf q}) \cdot {\bf q} + \Xi^\nu_u \left( {\bf e}^s({\bf q}) \cdot \hat{\bf k}_\nu \right) \left( {\bf q} \cdot \hat{\bf k}_\nu \right) \approx H_{nn} ({\bf k} ; {\bf q} s),
\label{ch5_Hep}
\end{equation}
\noindent where ${\bf e}^s(\vect{q})$ is the strain polarization vector.~Using Eq.~\eqref{ch5_Hep}, we derive the expressions for $H_{ep}$ in the deformation potential theory along high symmetry $\vect{q}$ directions for transverse and longitudinal acoustic phonons in terms of $\Xi_u^{\text{L}}$ and $\Xi_d^{\text{L}}$ \cite{Herring1956}. We then calculate the electron-phonon matrix elements $H_{nn} ({\bf k} ; {\bf q} s)$ using DFPT for the same phonon polarizations and directions. Finally, we extract the deformation potentials from the linear terms of polynomial fits to $H_{nn}(\vect{k}; \vect{q}s)$ versus $|\vect{q}|$ for all considered phonon polarizations and directions. Further details of the implementation of this approach are given in Appendix A.

We also calculate the deformation potentials of PbTe following the pioneering work of Van de Walle and Martin \cite{VandeWalle1986, VandeWalle1989, VandeWalle1987, VandeWalle1989a} on relative band shifts due to strain in semiconductors. We use this method as an alternative method to compute deformation potentials and validate those calculated from the DFPT approach. Furthermore, coupling this method with hybrid functionals, which give a better representation of electronic states near the band gap compared to LDA, allows us to properly include the effects of spin orbit coupling. The approach of Van de Walle and Martin has been applied to a wide range of semiconductors to capture band offsets at semiconductor interfaces and deformation potentials close to experiment \cite{VandeWalle1989}. A detailed discussion of the calculation of $\Xi_u^{\text{L}}$ and $\Xi_d^{\text{L}}$ following this approach is given in Appendix B. 

Our electronic band structure calculations, performed with the LDA exchange-correlation functional \cite{Hohenberg1964, Goedecker1996}, are carried out with the Hartwigsen-Goedecker-Hutter (HGH) norm-conserving pseudopotentials \cite{Hartwigsen1998} using the \textsc{Abinit} code \cite{Gonze2009,Gonze2016}. We use the Vienna \textit{ab initio} simulation package (\textsc{VASP}) \cite{Kresse1996, Kresse1996a} to perform electronic band structure calculations using the generalized gradient approximation (GGA) of Perdew, Burke, and Ernzerhof (PBE) \cite{Perdew1996, Perdew1997} and the screened Heyd-Scuseria-Ernzerhof (HSE03) hybrid functional \cite{Heyd2003, Heyd2004, Heyd2006, Paier2005}. In the PBE/HSE03 calculations, the basis set for the one-electron wave functions is constructed with the Projector Augmented Wave (PAW) method \cite{Blochl1994, Kresse1999}. We carry out the LDA calculations using the HGH pseudopotentials with the $6s^2 6p^2$ states of Pb and $5s^2 5p^4$ states of Te explicitly included in the valence states. For the PAW pseudopotentials, we also include the semicore $5d^{10}$ electronic states of Pb in the valence states since this leads to a better agreement of the band gap and the effective masses of PbTe with experiment~\footnote{The band gap and the parallel and perpendicular effective masses of the conduction band L valley of PbTe calculated using the HSE03 functional and the PAW pseudopotential that includes the $6s^2 6p^2$ valence states of Pb are $0.33$ eV, $0.3m_e$ and $0.035m_e$, respectively ($m_e$ is the free electron mass).}. We calculate the matrix elements $ \bra{u_{m\vect{k}+\vect{q}}} \partial_{\alpha b ,\vect{q}}v^{\text{KS}}  \ket{u_{n\vect{k}}}_{\text{uc}}$ directly from the density functional perturbation theory (DFPT) method \cite{Baroni2001, Giustino2017} as implemented in \textsc{Abinit} \cite{Gonze2009, Gonze1997, Gonze1997b}, using the LDA functional and the HGH pseudopotentials excluding SOI.

%%%%%%%%%%%%%%%%%%%
% Charge carrier effective masses

\section{III. Charge carrier effective masses}

\begin{figure}[!htb]
\begin{center}
\includegraphics[width=\graphsize\textwidth]{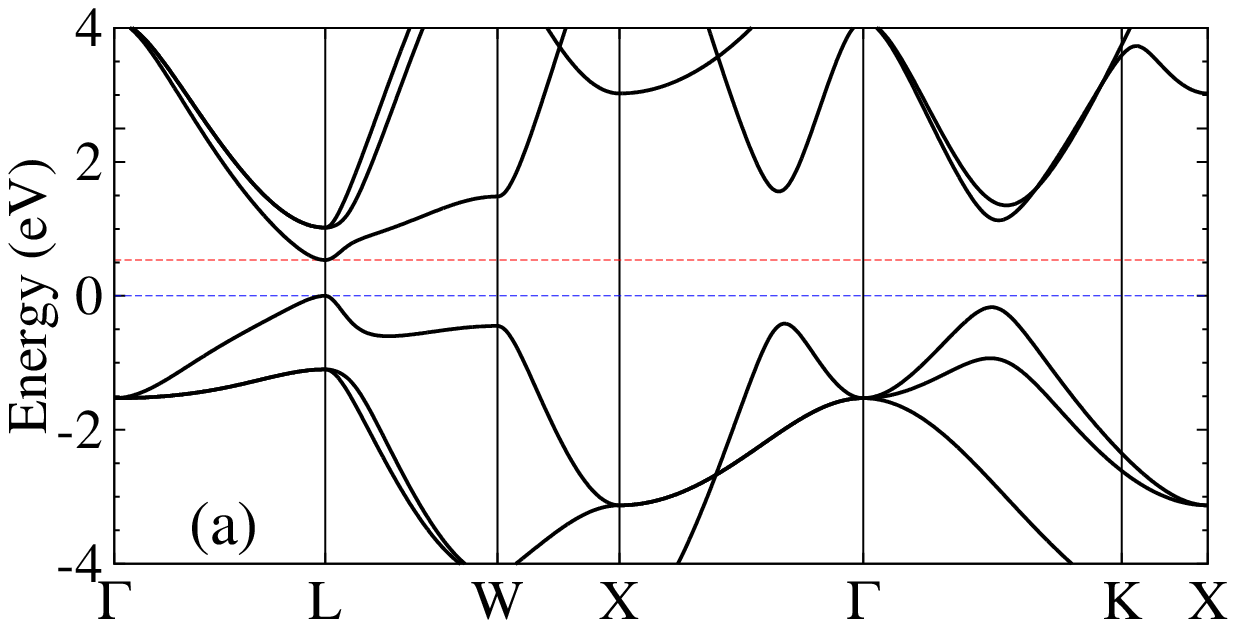}
\includegraphics[width=\graphsize\textwidth]{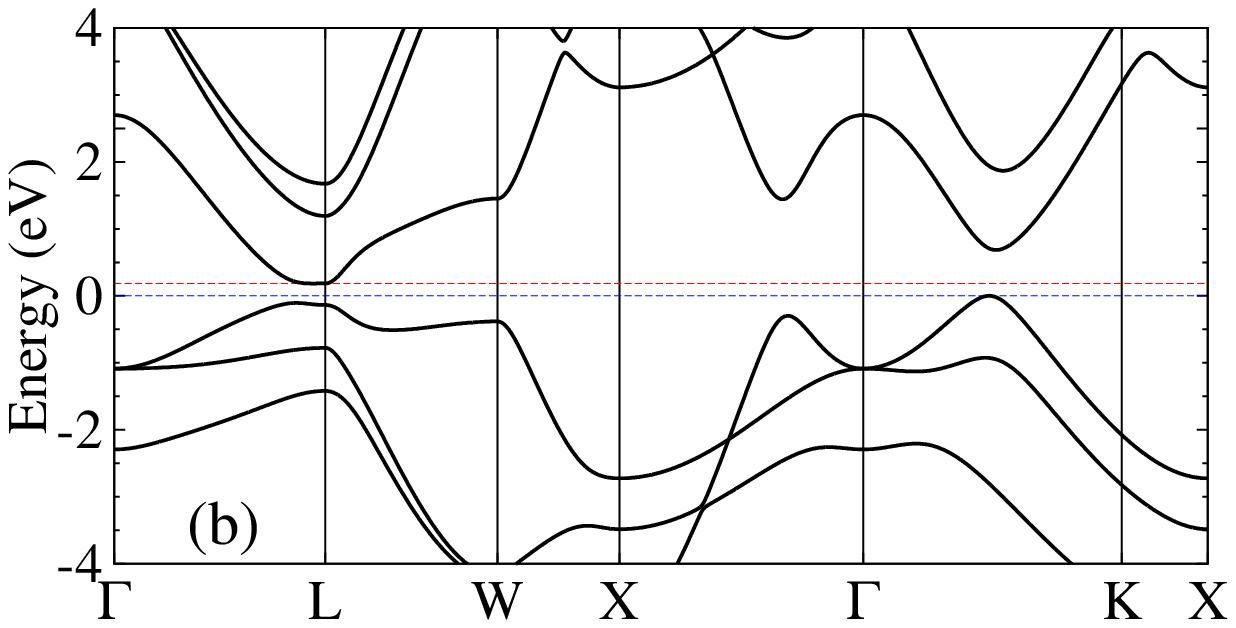}
\includegraphics[width=\graphsize\textwidth]{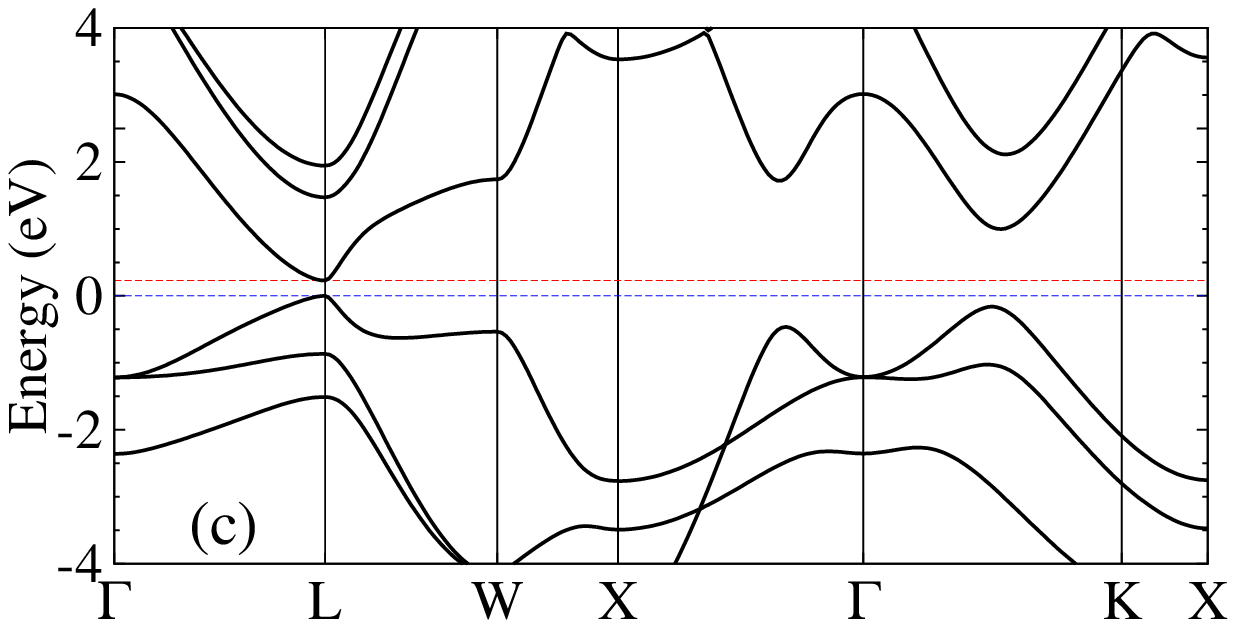}
\end{center}
\caption{Electronic band structure of PbTe, calculated using: (a) the LDA and HGH pseudopotentials excluding spin orbit interaction (SOI), (b) the LDA and HGH pseudopotentials including SOI, and (c) the HSE03 hybrid functional and PAW pseudopotentials including SOI. The energies of the valence band maximum (blue dashed line) and conduction band minimum (red dashed line) are also highlighted. The Fermi level at $0$ K has been set to $0$ eV.}
\label{ch5_elecbands_pbte}
\end{figure}

\begin{table*}[htb!]
\begin{center}
\begin{threeparttable}
\begin{tabularx}{\textwidth}{@{}l*8{>{\centering\arraybackslash}X}@{}}
\hline \hline
                                                 & LDA\ts{a}  & LDA\ts{a}   &PBE\ts{a}&HSE03\ts{a} & HSE03\ts{b}          & QS$GW$\ts{b}                & Exp. 1                & Exp. 2                   \\ [1ex] 
          & (exc. SOI) & (inc. SOI)    &(inc. SOI)&(inc. SOI)      &Ref.~\cite{Hummer2007}&Ref.~\cite{Svane2010a}&Ref.~\cite{Bauer2003}&Ref.~\cite{Dalven1974} \\ [1ex] \hline
E$_g$  (eV)                              & $0.54$      & $-0.32$~~  & 0.08       & 0.24            & 0.20                       & 0.29                      & 0.19\ts{c}          & 0.19\ts{c}             \\  [1ex] %0.081&0.237
$m_{\parallel}^{*,v}$/$m_e$ & 0.294        & 0.834          & 0.618      & 0.341          & 0.296                     & 0.338                    & 0.255                 & 0.310                     \\ [1ex]
$m_{\perp}^{*,v}$/$m_e$     & 0.023        & 0.622          & 0.184      & 0.030          & 0.029                     & 0.029                    & 0.024                  & 0.022                    \\ [1ex]
$m_{\parallel}^{*,c}$/$m_e$ & 0.216        & 5.295          & 0.362      & 0.246          & 0.223                     & 0.247                    & 0.210                  & 0.240                    \\ [1ex]
$m_{\perp}^{*,c}$/$m_e$     & 0.037        & 0.031          & 0.120      & 0.027          & 0.027                     & 0.027                    & 0.021                  & 0.024                    \\ [1ex]
\hline \hline
\end{tabularx}
\begin{tablenotes}
\item[a] {\footnotesize Calculated at the theoretically predicted 0 K lattice constant.}
\item[b] {\footnotesize Calculated at the low temperature experimental lattice constant.}
\item[c] {\footnotesize Increases to $\sim$0.3 eV at room temperature \cite{Tsang1971}.}
\end{tablenotes}
\end{threeparttable}
\end{center}
\caption{Electronic band gap (E$_g$) and valence and conduction band effective electron masses ($m^{*,v}$ and $m^{*,c}$, respectively) of PbTe calculated with the LDA, PBE, and HSE03 exchange-correlation functionals including/excluding spin orbit interaction (SOI), compared to the previous HSE03 and quasi-particle self-consistent $GW$ calculations and experiment. $m_e$ is the free electron mass.}
\label{table5_hse03}
\end{table*}

\begin{figure*}[!htp]
\begin{center}
\includegraphics[width=0.32\linewidth]{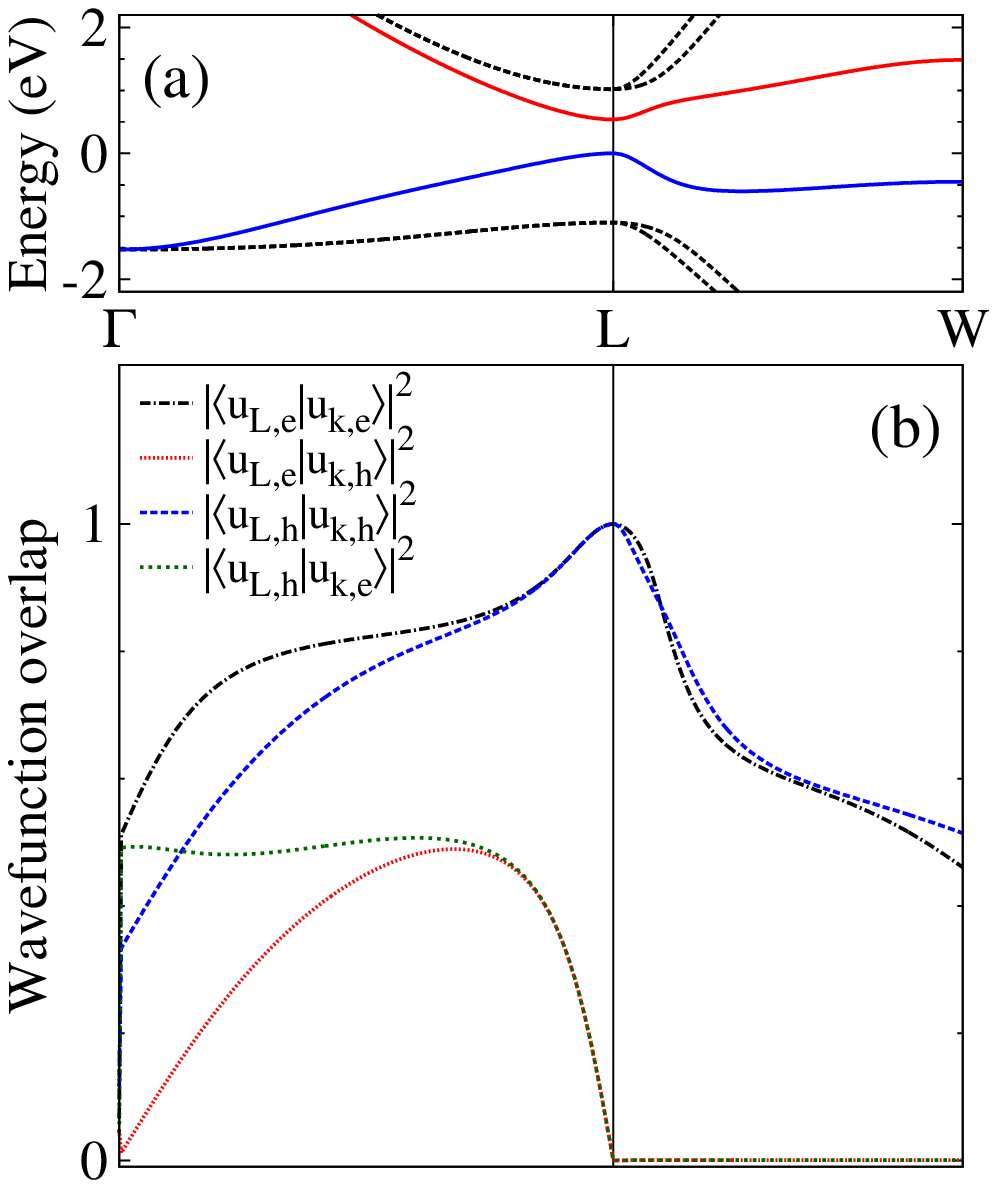}
\includegraphics[width=0.32\linewidth]{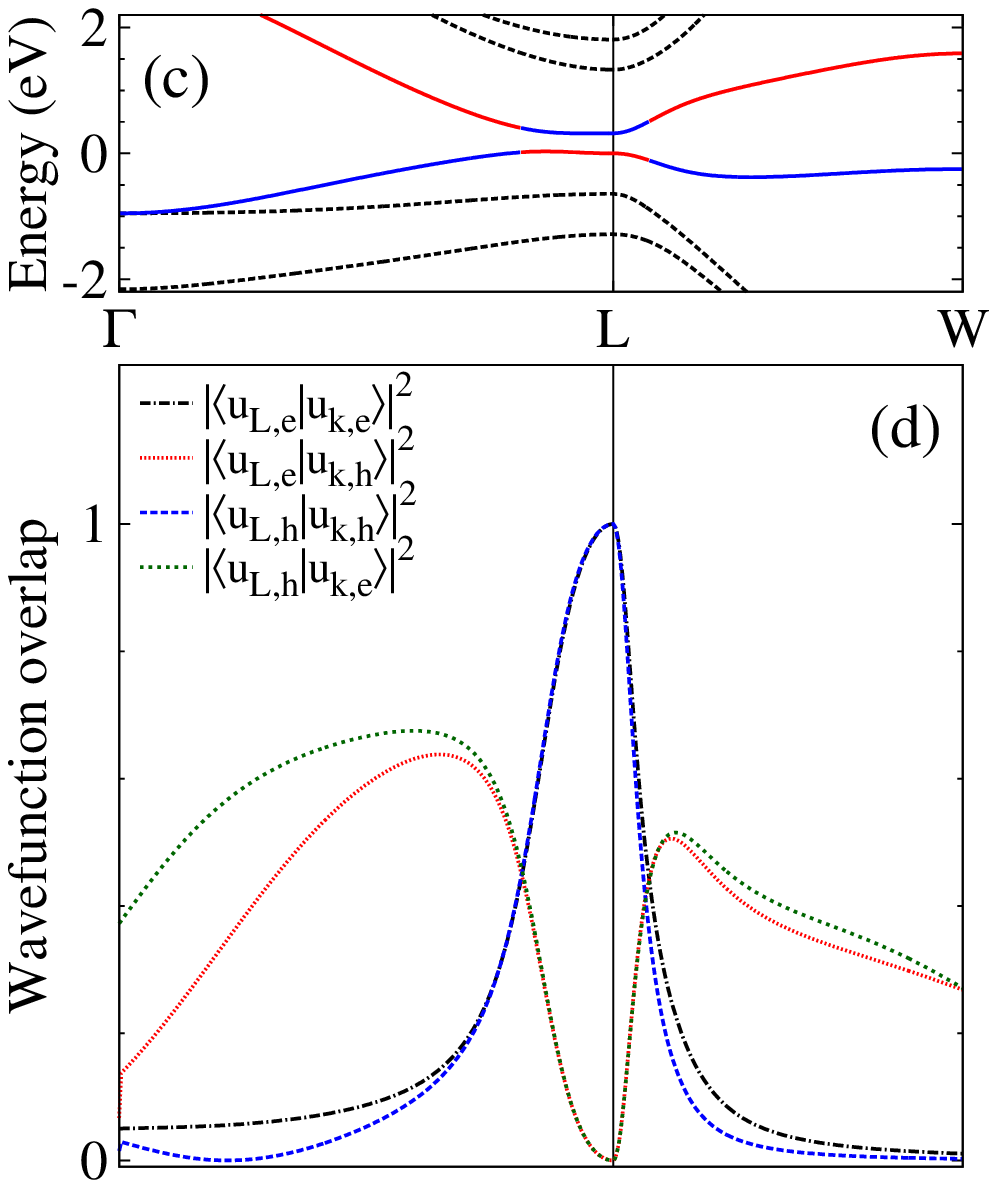}
\includegraphics[width=0.32\linewidth]{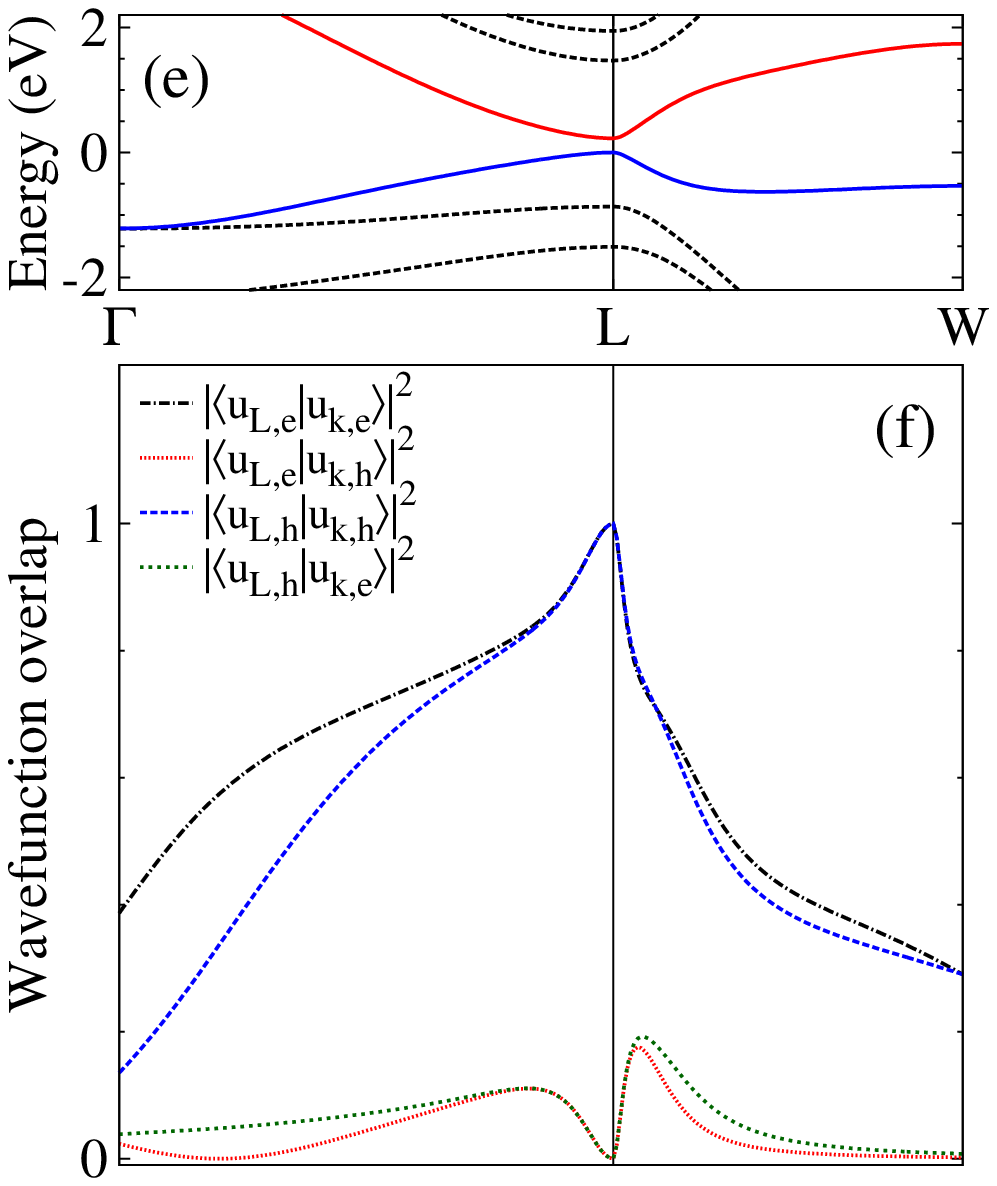}
\end{center}
\caption{Electronic band structure of PbTe near the Fermi level at $0$ K along the $\Gamma$-L-W line calculated using: (a) the LDA excluding spin orbit interaction (SOI), (c) the LDA including SOI, and (e) the HSE03 hybrid functional including SOI. A solid red (blue) line shows a state whose character mostly corresponds to that of the conduction (valence) band as deduced from wavefunction overlaps, see text for explanation. The overlaps $|\bra{u_{L}}\ket{u_{\vect{k}}}|^2$ of the periodic part of the wavefunction for the lowest conduction band (labeled as $e$) and the highest valence band (labeled as $h$) at the wavevector $\vect{k}$ with the wavefunction of one of these two bands at the L point calculated with: (b) the LDA excluding SOI, (d) the LDA including SOI, and (f) the HSE03 hybrid functional including SOI.}
\label{ch5_overlap}
\end{figure*}

Fig.~\ref{ch5_elecbands_pbte} shows the electronic band structure of PbTe calculated using: (a) the LDA and HGH pseudopotentials excluding SOI, (b) the LDA and HGH pseudopotentials including SOI, and (c) the HSE03 hybrid functional and PAW pseudopotentials including SOI \footnote{The LDA-HGH electronic structure calculations for a primitive unit cell were carried out using a four shifted $12\times 12\times 12$ $\vect{k}$-point grid for Brillouin zone sampling of electronic states and a $45$ Ha energy cutoff. The corresponding HSE03/PBE-PAW calculations were performed using a $8\times 8\times 8$ $\vect{k}$-point grid and a $18.4$ Ha energy cutoff.}. The combination of the LDA tendency to underestimate the band gap \cite{Perdew2009} and the effects of SOI results in an inverted band gap in PbTe. When relativistic effects are taken into account in the LDA functional, the previously degenerate 2\ts{nd} and 3\ts{rd} conduction and valence band states split, see Figs.~\ref{ch5_elecbands_pbte}(a) and (b). This spin orbit splitting causes the valence band maximum to be repelled upward, while the conduction band minimum is repelled downward \cite{Hummer2007, Wei1997}. The resulting band gap is underestimated to such a degree that the topmost valence band and bottommost conduction band become interchanged and mix heavily near the L point \cite{Hummer2007, Wei1997}, forming  a ``negative band gap'' \cite{Svane2010a}. Furthermore, the $\Sigma$ valley (along the $\Gamma$-K direction) becomes the highest valence band, forming an indirect band gap with the lowest unoccupied energy state at L, see Fig.~\ref{ch5_elecbands_pbte}(b), at odds with experimental observations \cite{Bauer2003, Dalven1974, Tsang1971}. 

In order to precisely determine whether we have a positive or negative band gap, we compute the overlap of the periodic parts of the conduction and valence band wavefunctions at the L point with the wavefunctions of the same bands at wavevectors away from L. We consider only the 1\ts{st} electronic states on either side of the Fermi level, which corresponds to the highest occupied state at 0 K, and is set to 0 eV, see Fig.~\ref{ch5_elecbands_pbte}. We label the periodic part of the wavefunction of the state below the Fermi level $\ket{u_{\vect{k},\text{h}}}$, and the periodic part of the wavefunction of the state above the Fermi level $\ket{u_{\vect{k},\text{e}}}$. We define the valence band state at L as the band state that satisfies $|\bra{u_{L,\text{h}}}\ket{u_{\vect{k},\text{h}}}| > |\bra{u_{L,\text{h}}}\ket{u_{\vect{k},\text{e}}}|$ for wavevectors $\vect{k}$ sufficiently far from L. The conduction band state at L is similarly defined as the band state that satisfies $|\bra{u_{L,\text{e}}}\ket{u_{\vect{k},\text{e}}}| > |\bra{u_{L,\text{e}}}\ket{u_{\vect{k},\text{h}}}|$ for wavevectors $\vect{k}$ sufficiently far from L. If the character for the conduction and valence band states changes as a function of $\vect{k}$, the values of $|\bra{u_{L,\text{h}}}\ket{u_{\vect{k},\text{h}}}|$ and $|\bra{u_{L,\text{e}}}\ket{u_{\vect{k},\text{e}}}|$ will become substantially smaller than those of $|\bra{u_{L,\text{h}}}\ket{u_{\vect{k},\text{e}}}|$ and $|\bra{u_{L,\text{e}}}\ket{u_{\vect{k},\text{h}}}|$ respectively, as the $\vect{k}$-vector moves away from L, see Fig.~\ref{ch5_overlap}(d). We choose the wavevector $\vect{k}$ to be along the $\Gamma$-L-W line, which corresponds to the directions along which we extract the values of $m_{\parallel}^{*}$ ($\text{L}\rightarrow \Gamma$) and $m_{\perp}^{*}$ ($\text{L}\rightarrow \text{W}$). We plot these wavefunction overlaps in Fig.~\ref{ch5_overlap} for: (b) the LDA and HGH pseudopotentials excluding SOI, (d) the LDA and HGH pseudopotentials including SOI, and (f) the HSE03 hybrid functional and PAW pseudopotentials including SOI.

By computing the described wavefunction overlaps, we confirm that the inclusion of SOI within the LDA-HGH level of theory results in a negative band gap. Fig.~\ref{ch5_overlap}(d) illustrates that the band state at L above the Fermi level overlaps more strongly with the band states below the Fermi level for wavevectors away from L. Therefore, $\ket{u_{L,\text{e}}}$ mainly exhibits the character of the valence band. Conversely, the band state at L below the Fermi level overlaps more strongly with the band states above the Fermi level for wavevectors away from L, which indicates that $\ket{u_{L,\text{h}}}$ has the character of the conduction band. This analysis confirms the change of the conduction and valence band states character as the wavevector $\vect{k}$ varies, see Fig.~\ref{ch5_overlap}(c), and the presence of an inverted band gap. As a result, effective masses calculated with the LDA functional including SOI \footnote{Effective masses were calculated by fitting Eq.~\eqref{ch5_parabands} over a $\vect{k}$-vector range of approximately $1/20$ of the distance of the $\Gamma$-L and L-W lines respectively.} are in poor agreement with experiment, see Table~\ref{table5_hse03}.

In contrast, the LDA-HGH level of theory excluding SOI correctly represents the character of the conduction and valence band states near the gap in PbTe despite the band gap overestimation, see Fig.~\ref{ch5_overlap}(b). The band state at L below (above) the Fermi level overlaps more strongly with the band states above (below) the Fermi level for wavevectors away from L, and thus it mainly exhibits the character of the valence (conduction) band. This confirms the correct ordering of states at L and a positive band gap when SOI is excluded, and verifies that our effective electron masses are extracted from physically correct states. Indeed, the effective electron masses of PbTe calculated with the LDA excluding SOI are very close to experiments, see Table~\ref{table5_hse03}. We thus anticipate that the LDA excluding SOI may reasonably well describe electronic transport properties of PbTe. We note that the energy separation of the 1\ts{st} and 2\ts{nd} conduction band states without SOI is considerably smaller than in the case including SOI, which may lead to some inaccuracies in the electron-phonon matrix element calculations.

In order to more accurately describe the energy differences and curvature of the states near the band gap of PbTe, we also carried out hybrid functional calculations. The effectiveness of screened hybrid functionals \cite{Heyd2003} in capturing structural, elastic, and electronic properties has previously been shown for a wide range of semiconductors \cite{Paier2006, Heyd2005, Paier2006a}. In particular, the documented ability of the HSE03 functional \cite{Heyd2003, Heyd2004, Heyd2006} to correctly reproduce the band gap and correct ordering of states with the inclusion of SOI in lead chalcogenides \cite{Hummer2007} is the main motivation for its use in this work. We show the electronic band structure of PbTe calculated with the HSE03 functional including SOI in Fig.~\ref{ch5_elecbands_pbte}(c). We note that the PBE functional including SOI, which was used to construct the HSE03 functional, yields a very narrow positive band gap and poor effective masses compared to experiment, see Table \ref{table5_hse03}.

The HSE03 functional including SOI reproduces the band gap and effective masses of PbTe to excellent agreement with experiment and previous hybrid functional \cite{Hummer2007} and quasi-particle self-consistent $GW$ (QS$GW$) \cite{Svane2010a} calculations, see Table~\ref{table5_hse03}. This is due to the fact that the HSE03 functional obtains the correct ordering of states near the band gap, which we confirmed by computing the previously defined wavefunction overlap, see Figs.~\ref{ch5_overlap}(e)~and~(f). Thus, there is no entanglement of the conduction and valence bands in HSE03 calculations as obtained using the LDA with SOI \cite{Hummer2007}. Furthermore, the valence band maximum is found at L, and not along the $\Gamma$-K direction, as was the case for the LDA including SOI. The HSE03 functional also captures a much larger energy separation of the 1\ts{st} and 2\ts{nd} conduction band states compared to the LDA without SOI. This potentially resolves any issues whereby incorrect hybridization with the 2\ts{nd} conduction band might compromise the calculation of the electron-phonon matrix elements near the conduction band minimum. 

\begin{table}[htb!]
\begin{center}
\begin{threeparttable}
\begin{tabularx}{0.475\textwidth}{@{}l*2{>{\centering\arraybackslash}X}@{}}
\hline \hline
                                                                                            &$dE_g/dP$ (inc. SOI)           &$dE_g/dP$ (exc. SOI)          \\ [1ex] 
                                                                                            & ($\times10^{-6}$ eV/Bar)   & ($\times10^{-6}$ eV/Bar)  \\ [1ex] \hline
LDA                                                                                      &$+2.1$                                 & $-4.6$                                 \\ [1ex] 
PBE                                                                                      &$-4.9$                                  & $-$                                  \\ [1ex] 
HSE03                                                                                  &$-7.0$                                  & $-$                                       \\ [1ex] \hline
Exp.~\cite{Zasavitskii2004, Schluter1975} & $-7.4$                   &       \\  [1ex] 

Exp.~\cite{Ravindra1980} &  $-7.5$                   &           \\  [1ex]\hline \hline
\end{tabularx}
\end{threeparttable}
\end{center}
\caption{Hydrostatic pressure gap coefficient of PbTe calculated with the LDA, PBE, and HSE03 functionals with and without spin orbit interaction (SOI), compared to measurements at 4 K \cite{Zasavitskii2004}, 77 K \cite{Schluter1975}, and 300 K \cite{Ravindra1980}.}
\label{table5_dedg}
\end{table}

We further confirm the accurate description of the electronic band structure of PbTe using the HSE03 functional by comparing the calculated hydrostatic pressure gap coefficient with experiment. The hydrostatic pressure gap coefficient is defined as the change in the electronic band gap with respect to pressure, $dE_g/dP$ \footnote{We simulated pressure by setting the lattice constant to $0.999\times a_0$, where $a_0$ is its theoretically predicted equilibrium value}. The value of $dE_g/dP$ calculated with the HSE03 functional compares very well with experimental data~\cite{Zasavitskii2004, Schluter1975, Ravindra1980}, and captures a decreasing band gap with pressure, see Table~\ref{table5_dedg}. Thus, we expect that the hybrid functional will accurately describe changes in the electronic band structure due to strain or pressure, of particular importance when computing deformation potentials.

The LDA functional excluding SOI also gives the correct sign of $dE_g/dP$ since it correctly captures the sign of the band gap. The calculated value of $dE_g/dP$  with the LDA excluding SOI is in fair agreement with experiment, see Table~\ref{table5_dedg}. This indicates that the DFT-LDA level of theory excluding SOI may give fairly accurate values of the deformation potentials of PbTe. However, the agreement with experiment is not as good as that of the HSE03 functional, possibly due to the small energy separation of the 1\ts{st} and 2\ts{nd} conduction band states. In contrast, the calculated value with the LDA including SOI yields an increasing band gap with pressure due to the inverted ordering of the conduction and valence bands and heavy mixing of states near the band edges \cite{Hummer2007}. This finding further suggests that the LDA including SOI may not reliably describe the deformation potentials of PbTe.

%%%%%%%%%%%%%%%%%%%
% Deformation potentials

\section{IV. Deformation potentials of PbTe}

We calculate the acoustic deformation potentials of PbTe using all the exchange-correlation functionals considered so far. We expect that the LDA functional excluding SOI and the HSE03 functional including SOI will give reasonable values since they both give a good description of the electronic band structure of PbTe. To compute deformation potentials using the LDA excluding SOI, we use both DFPT and the relative band shifts approach proposed by Van de Walle and Martin \cite{VandeWalle1986, VandeWalle1989, VandeWalle1987, VandeWalle1989a}. For the HSE03, PBE and LDA functionals including SOI, we use the approach of Van de Walle and Martin \cite{VandeWalle1986, VandeWalle1989, VandeWalle1987, VandeWalle1989a}.

\begin{table}[htb!]
\begin{center}
\begin{threeparttable}
\begin{tabularx}{0.49\textwidth}{@{}l*4{>{\centering\arraybackslash}X}@{}}
\hline  \hline  
                                                          & DFPT          & Band shift     & Previous calculation                           & Experiment                                              \\ [1ex]\hline
$\Xi_{u}^{\text{L}}$ ($\text{eV}$) & $~~7.0$  & $ 8.0$         & $~~8.3$ \ts{\cite{Ferreira1965}}  & $3, 4.5$  \ts{\cite{Ravich1971a}}         \\ [1ex] 
$\Xi_{d}^{\text{L}}$ ($\text{eV}$) & $~~0.4$      & $ 0.5$         & $-4.4$ \ts{\cite{Ferreira1965}}      & $17-22$\ts{a} \ts{\cite{Ravich1971a}} \\ [1ex]\hline \hline
\end{tabularx}
\begin{tablenotes}
\item[a] {\footnotesize Obtained by fitting a model to electronic transport measurements.}
\end{tablenotes}
\end{threeparttable}
\end{center}
\caption{Deformation potentials of PbTe calculated using density functional perturbation theory (DFPT), and the relative energy shift of the L valley under strain (see text for explanation), using the LDA and HGH pseudopotentials excluding spin orbit interaction. Our values are compared to those from previous calculations and experiment.}
\label{table5_defpot_pbte}
\end{table}

We first compare the calculated values of the deformation potentials of PbTe, obtained using DFPT with the LDA functional excluding SOI, with those from experiments~\cite{Lalonde2011, Ravich1971a, Ravich1971}, see Table~\ref{table5_defpot_pbte}. The computed uniaxial deformation potential of PbTe, $\Xi_u^{\text{L}}=7.0$ eV, agrees reasonably well with the experimental values. The value of $\Xi_u^{\text{L}}=3$ eV was obtained from piezoresistance measurements, whereas the value of $\Xi_u^{\text{L}}=4.5$ eV was obtained from the ultrasonic technique, see Refs.~\cite{Ravich1971a, Ravich1971} and references therein. In contrast, our calculated value of the dilatation deformation potential of PbTe, $\Xi_d^{\text{L}}=0.4$ eV, is considerably lower than the corresponding values deduced by fitting a model to electronic transport measurements, $\Xi_d^{\text{L}}\sim 17-22$ eV~\cite{Lalonde2011, Ravich1971a, Ravich1971}. The large discrepancy between our calculated values of $\Xi_d^{\text{L}}$ and those evaluated from experiments will be discussed later in more detail.

Our value of $\Xi_u^{\text{L}}=7.0$ eV calculated using DFPT with the LDA excluding SOI agrees fairly well with that of an early empirical pseudopotential calculation, $\Xi_u^{\text{L}}=8.3$ eV \cite{Ferreira1965}, see Table~\ref{table5_defpot_pbte}. In contrast, our calculated value of $\Xi_d^{\text{L}}=0.4$ eV differs from the value of $\Xi_d^{\text{L}}=-4.4$ eV obtained in the same work. However, we note that these previous values were obtained with an empirical, non-self-consistent augmented plane wave method \cite{Ferreira1965}. Thus, we do not expect quantitative agreement with these results. Interestingly, both our and the previous computed values for $\Xi_d^{\text{L}}$ differ significantly from the values estimated from electronic transport measurements.

Next we compare the values of deformation potentials calculated using DFPT and the approach of Van de Walle and Martin for the LDA-HGH level of theory excluding SOI, also shown in Table~\ref{table5_defpot_pbte}. The value of $\Xi_u^{\text{L}}=8$~eV calculated from the strain induced splitting of the L valleys agrees well with the value of $\Xi_u^{\text{L}}=7$~eV from DFPT. Furthermore, the calculated value of $\Xi_d^{\text{L}}=0.5$~eV using the relative band shift method for the LDA functional without SOI is close to that calculated using DFPT, $\Xi_d^{\text{L}}=0.4$~eV. We thus find an overall good agreement between our calculated values of deformation potentials using the two methods, and ascribe the small differences between the two methods to the various numerical approximations described in detail in Appendices A and B. 

\begin{table}[htb!]
\begin{center}
\begin{threeparttable}
\begin{tabularx}{0.49\textwidth}{@{}l*4{>{\centering\arraybackslash}X}@{}}
\hline \hline
                                                            & LDA           & LDA            & PBE           & HSE03          \\ 
                                                            & (exc. SOI) & (inc. SOI)  & (inc. SOI)   & (inc. SOI)     \\ [1ex]\hline
$\Xi_{u}^{\text{L}}$ ($\text{eV})$   & $8.0$      & $~~7.7$ & $~~7.3$ & $5.5$         \\ [1ex] %& $5.66$     & $6.81$ 
$\Xi_{d}^{\text{L}}$ ($\text{eV})$   & $0.5$      & $-1.6$     & $-0.2$     & $0.8$         \\ [1ex]\hline\hline %& $0.31$     & $0.33$ 
\end{tabularx}
\end{threeparttable}
\end{center}
\caption{Deformation potentials of PbTe calculated from the relative energy shifts of the L valley under strain using the LDA, PBE and HSE03 exchange-correlation functionals including/excluding spin orbit interaction (SOI).}
\label{table5_defpotexc}
\end{table}

The deformation potentials of PbTe calculated with the HSE03 functional are fairly close to those of the LDA excluding SOI, see Tables~\ref{table5_defpot_pbte} and \ref{table5_defpotexc}. The value of $\Xi_u^{\text{L}}=5.5$~eV calculated with the HSE03 functional is in better agreement with the experimental value of $\Xi_u^{\text{L}}=4.5$~eV than those of the LDA excluding SOI. We note that if we exclude the $5d^{10}$ electronic states of Pb from the valence states of PAW pseudopotentials, we obtain $\Xi_u^{\text{L}}=6.6$~eV, in better agreement with the LDA-HGH result of $\Xi_u^{\text{L}}=8.0$~eV obtained using the Van de Walle and Martin approach. This suggests that the difference between these two functionals may be due to the inaccurate description of the energy separation of the 1\ts{st} and 2\ts{nd} conduction band states in the LDA excluding SOI. The HSE03 functional yields a value of $\Xi_d^{\text{L}}=0.8$~eV, consistent with those calculated with the LDA excluding SOI. Consequently, the values of $\Xi_d^{\text{L}}$ calculated with both the HSE03 including SOI and the LDA excluding SOI are substantially smaller than the values of $\Xi_d^{\text{L}}=17-22$~eV estimated from electronic transport measurements.

The consistency of our dilatation deformation potential values calculated using different approaches suggests that the discrepancy with the values evaluated from transport measurements is unlikely to be due to an error in our implementation of either approach. Furthermore, our effective masses and uniaxial deformation potential agree very well with previous measurements. We note that the values of dilatation deformation potentials extracted from experiments depend on the chosen transport model and the parameter values used. In extracting the deformation potentials of PbTe, Ravich \textit{et al.}~\cite{Ravich1971a} assumed that the electronic mobility as a function of carrier concentration at 77 K is limited by three scattering mechanisms: impurity scattering, polar optical scattering, and acoustic phonon scattering. Using experimental parameters for the former two mechanisms, the electron mobility limited by these was subtracted from the experimental values. An effective deformation potential was then obtained by fitting a relatively simple model to the remaining mobility. On the other hand, LaLonde \textit{et al.}~\cite{Lalonde2011} entirely attributed electron-phonon scattering to acoustic mode phonons, and fitted their model to electronic transport measurements for a range of temperatures and carrier concentrations. Based on all these considerations, we conclude that the dilatation deformation potential is indeed of the order of $\Xi_d^{\text{L}}\sim 0.5$ eV as our calculations show, and thus much weaker than previously believed.

Interestingly, the deformation potential values calculated using the the approach of Van de Walle and Martin using different exchange-correlation functionals are in fairly good agreement with each other, even when their band structures are not correctly described, see Table \ref{table5_defpotexc}. When SOI is accounted for at the LDA level of theory, we calculate $\Xi_u^{\text{L}}=7.7$ eV from the strain induced L valley splitting. This value is close to the values calculated using the LDA functional excluding SOI and the HSE03 functional including SOI. Furthermore, the value of $\Xi_d^{\text{L}}=-1.6$ eV obtained using the LDA including SOI is not dramatically different from those calculated using the LDA excluding SOI and the HSE03 functionals (but it has a different sign). The deformation potential values obtained using the PBE functional excluding SOI are also consistent with all these values.

The analysis above indicates that obtaining the correct character of the states near the band gap may not be critically important for determining the correct order of magnitude of the deformation potentials of PbTe. This may be due to the fact that the conduction and valence bands at L have a mirror symmetry, so that their inversion does not have a large influence on the deformation potentials. Furthermore, the values of dilatation deformation potentials are very small and their change of sign depending on the band edge states character may not affect charge transport much. However, reliable transport calculations require accurate effective masses in addition to accurate deformation potentials. Only those exchange-correlation functionals which capture all these quantities correctly will be able to properly describe electronic transport. Our analysis shows that the LDA excluding SOI and the HSE03 including SOI may provide a sufficiently accurate representation of the electronic transport properties of PbTe.
 
%%%%%%%%%%%%%%%%%%%
% Conclusion

\section{VII. Conclusion}

The local density approximation (LDA) excluding spin orbit coupling and the HSE03 hybrid functional capture a positive band gap and physically correct conduction and valence band states near the band gap in PbTe. We calculate the uniaxial and dilatation acoustic deformation potentials of the L valley conduction band of PbTe using density functional perturbation theory and the LDA excluding spin orbit coupling, and find values of $\Xi_{u}^{\text{L}}=7.0$~eV and $\Xi_{d}^{\text{L}}=0.4$~eV, respectively. We validate these values using an alternative approach to compute deformation potentials from the relative shifts of the L valleys under strain. Coupling this approach with the HSE03 hybrid functional also allowed us to accurately include the effects of spin orbit interaction, yielding values of $\Xi_{u}^{\text{L}}=5.5$~eV and $\Xi_{d}^{\text{L}}=0.8$~eV. Our results show that a particular choice of an exchange-correlation functional is not critically important for calculating the effective masses and deformation potentials of PbTe, provided the functional captures physically correct states near the band edges. Our dilatation deformation potential values of PbTe are much lower than those obtained by fitting electronic transport measurements, suggesting that intravalley acoustic phonon scattering is considerably weaker than assumed in prior studies. 

\section{Acknowledgements}

We thank Jiang Cao and Jos\'e Daniel Querales-Flores for useful discussions.~This work was supported by Science Foundation Ireland (SFI) and the Marie-Curie Action COFUND under Starting Investigator Research Grant 11/SIRG/E2113, and by SFI under Investigators Programme 15/IA/3160. We acknowledge the Irish Centre for High-End Computing (ICHEC) for the provision of computational facilities. 

%%%%%%%%%%%%%%%%%%%
% Appendices

\section{Appendix A: Deformation potentials from density functional perturbation theory}

We calculate deformation potentials due to long-wavelength acoustic phonons \cite{Bardeen1950,Herring1956} using DFPT and following the approach of Fischetti and Laux \cite{Fischetti} and Murphy-Armando and Fahy \cite{Murphy-Armando2008}. We note that DFPT is a reciprocal space perturbative approach to calculate electron-phonon matrix elements, which correspond to acoustic deformation potentials only in the limit of ${\bf q}\rightarrow 0$. The detailed steps of this procedure are given below.

\subsection{\textit{Deformation potentials}}

We first obtain the expressions for the acoustic deformation potentials along high symmetry $\vect{q}$ directions in terms of $\Xi_u^{\text{L}}$ and $\Xi_d^{\text{L}}$, using the deformation potential Hamiltonian $H_{ep}$ given by Eq.~(\ref{ch5_Hep}). Since PbTe is a cubic material, we have five high symmetry $\vect{q}$ directions to consider: $[010]$, $[0\bar{1}1]$, $[111]$, $[011]$, and $[1\bar{1}1]$. We write the expressions for the deformation potentials $H_{ep}/q$ of the [111] L valley by inserting the appropriate strain polarization vector $\vect{e}^{s}(\vect{q})$ for a given $\vect{q}$ vector into Eq.~(\ref{ch5_Hep}). The expressions for $H_{ep}/q$ for the five $\vect{q}$ directions and the associated transverse and longitudinal acoustic phonons are given in Table~\ref{table5_directions}, which corresponds to Table IV of Herring and Vogt \cite{Herring1956}. We then calculate all listed $H_{ep}/q$, using Eq.~(\ref{ch5_Hep}) and the related electron-phonon matrix elements computed with DFPT, as described below.

\bgroup
\def\arraystretch{1.25}
\begin{table}[htb!]
\begin{center}
\begin{threeparttable}
\begin{tabularx}{0.49\textwidth}{@{}l*3{>{\centering\arraybackslash}X}@{}}
\hline  \hline  
$\vect{q}$      & TA$_1$                                    & TA$_2$                                      & LA                                                \\ [1ex]
\hline
$[010]$           & $\frac{1}{3}\Xi_u^{\text{L}}$              & $\frac{1}{3}\Xi_u^{\text{L}}$               & $\Xi_d^{\text{L}} + \frac{1}{3}\Xi_u^{\text{L}}$ \\ [1ex] 
$[0\bar{1}1]$ & $0$                                                         & $0$                                                           & $\Xi_d^{\text{L}}$                                                  \\ [1ex] 
$[111]$           & $0$                                                         & $0$                                                           & $\Xi_d^{\text{L}} + \Xi_u^{\text{L}}$                  \\ [1ex] 
$[110]$           & $\sqrt{\frac{2}{9}}\Xi_u^{\text{L}}$   & $0$                                                          & $\Xi_d^{\text{L}} + \frac{2}{3}\Xi_u^{\text{L}}$ \\ [1ex]
$[1\bar{1}1]$ & $\sqrt{\frac{2}{27}}\Xi_u^{\text{L}}$ & $-\sqrt{\frac{2}{81}}\Xi_u^{\text{L}}$  & $\Xi_d^{\text{L}} + \frac{1}{9}\Xi_u^{\text{L}}$ \\ [1ex] 
\hline \hline
\end{tabularx}
\end{threeparttable}
\end{center}
\caption{The expressions for $H_{ep}/q$ obtained from Eq.~(\ref{ch5_Hep}) for the transverse and longitudinal acoustic phonons along the five high symmetry directions of $\vect{q}$ for the [111] L valley of a cubic material. $\Xi_u^{\text{L}}$ and  $\Xi_d^{\text{L}}$ represent uniaxial and dilatation deformation potentials, respectively. Reproduced from Table~IV of Ref.~\cite{Herring1956}.}
\label{table5_directions}
\end{table}
\egroup

\subsection{\textit{Matrix element calculation details}}

We compute the electron-phonon matrix elements given by Eq.~\eqref{ch2_elphmatelement} for the transverse and longitudinal acoustic phonons along the five high symmetry $\vect{q}$ directions using the DFPT method \cite{Baroni2001, Giustino2017}, as implemented in \textsc{Abinit} \cite{Gonze2009, Gonze1997, Gonze1997b}. We used the LDA exchange-correlation functional excluding SOI and the HGH pseudopotentials. In this work, we are only interested in intravalley acoustic phonon scattering involving the conduction band L valleys. Thus, we calculate the electron-phonon matrix elements coupling states $\text{L}$ and $\text{L}+\vect{q}$ (labeled as $\text{L}\rightarrow\text{L}+\vect{q}$), as well as the matrix elements coupling states $\text{L}-\vect{q}/2$ and $\text{L}+\vect{q}/2$ (labeled as $\text{L}-\vect{q}/2\rightarrow\text{L}+\vect{q}/2$). To extract the deformation potential values as $\vect{q}\rightarrow 0$, we calculate the electron-phonon matrix elements for a range of $\vect{q}$ along the five symmetry directions, including as short $\vect{q}$ as the computational cost allows. The phonon wavevectors used in our DFPT calculations are the shortest wavevectors that are commensurate with the $\vect{k}$-point grids for electronic states ranging from $8\times8\times8$ to $48\times48\times48$. We use an energy cutoff of 20 Ha for plane waves. 

\subsection{\textit{Subtracting Fr\"ohlich interaction}}

PbTe is a polar semiconductor and thus some portion of the electron-phonon matrix element will be due to the interaction of electrons with the electric field generated by oppositely charged atoms moving out of phase with each other~\cite{Frohlich1954}. This Fr\"ohlich contribution to electron-phonon coupling will be large for longitudinal optical (LO) phonons, but it may also be non-negligible for acoustic phonons since their eigenvectors may contain a small LO component. To obtain deformation potentials due to acoustic vibrations only, we subtract the Fr\"ohlich contribution from the electron-phonon matrix elements calculated using DFPT~\cite{Vogl1976, Sjakste2015, Verdi2015}. To be consistent with our definition of the electron-phonon matrix element given in Eq.~\eqref{ch2_elphmatelement}, we define those due to the Fr\"ohlich interaction as~\cite{Giustino2017,Verdi2015}:
\begin{eqnarray}
H^{F}_{mn}(\vect{k}; \vect{q}s) = i\frac{e^2}{\Omega \varepsilon_0} \bra{u_{m\vect{k}+\vect{q}}}\ket{u_{n\vect{k}}}_{\text{uc}} \sum_{b} \left( \frac{m_{c}}{m_{b}} \right)^{\frac{1}{2}} \times \nonumber\\
\sum_{\vect{G}}\frac{(\vect{q}+\vect{G})\cdot\vect{Z}^*_{b}\cdot\vect{e}^s_b(\vect{q})}{(\vect{q}+\vect{G})\cdot \vect{\varepsilon_{\infty}} \cdot (\vect{q}+\vect{G})}e^{-i(\vect{q}+\vect{G})\vect{\tau}_b}e^{-|\vect{q}+\vect{G}|^2/4\alpha},
\label{ch5_frohlich}
\end{eqnarray}
\noindent where $\Omega$ is the volume of the primitive cell, and $\varepsilon_0$ is the permittivity of free space. $\vect{G}$ is the reciprocal lattice vector and $\vect{\tau}_b$ is the position of atom $b$. Finally, $\alpha$ is the convergence parameter for the summation over $\vect{G}$ in Eq.~\eqref{ch5_frohlich}, taken as $\alpha=5(2\pi/a)^2$, where $a$ is the lattice constant \cite{Sjakste2015}. $\vect{Z}^*_{b}$ and $\vect{\varepsilon_{\infty}}$ are Born effective charge and high-frequency dielectric tensors, respectively. They are diagonal and isotropic for cubic materials, and we calculated their values using DFPT and the LDA excluding SOI, yielding $Z^*_{\text {Pb}}=6.35$, $Z^*_{\text {Te}}=-6.35$ and $\varepsilon_{\infty}=34.85$.

\begin{figure*}[!htp]
\begin{center}
\includegraphics[width=0.31\linewidth]{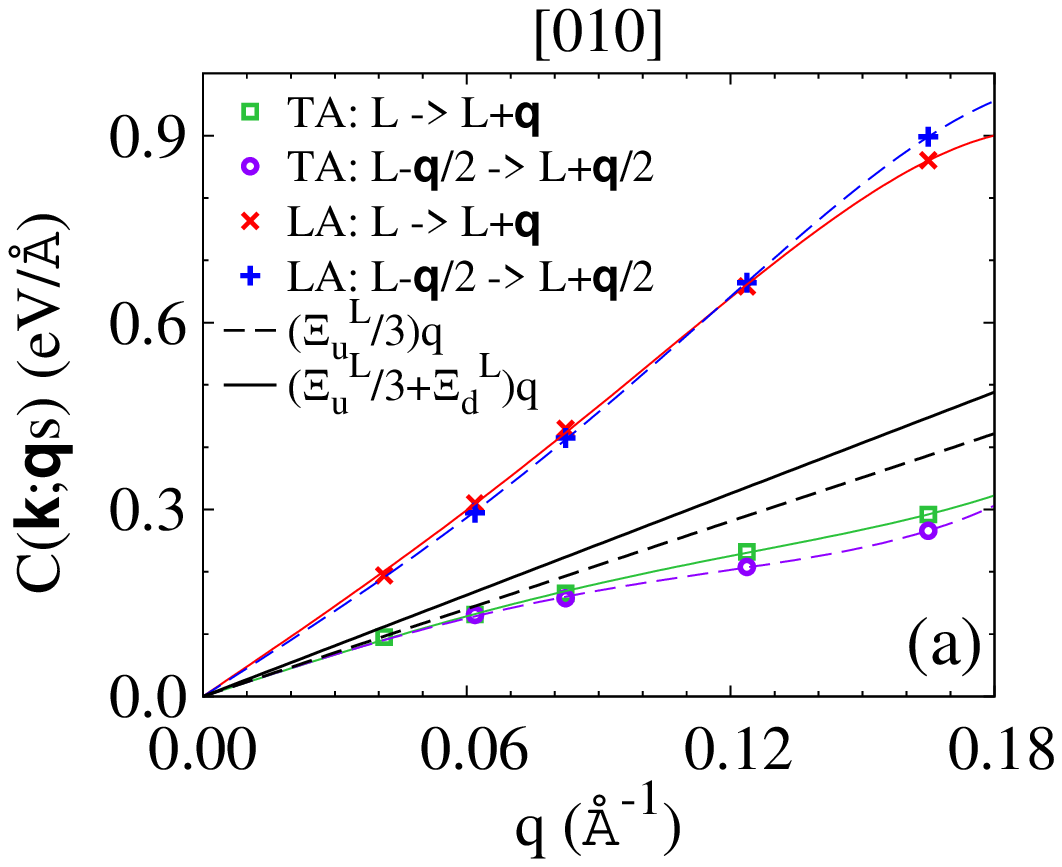} \hspace{3mm}
\includegraphics[width=0.31\linewidth]{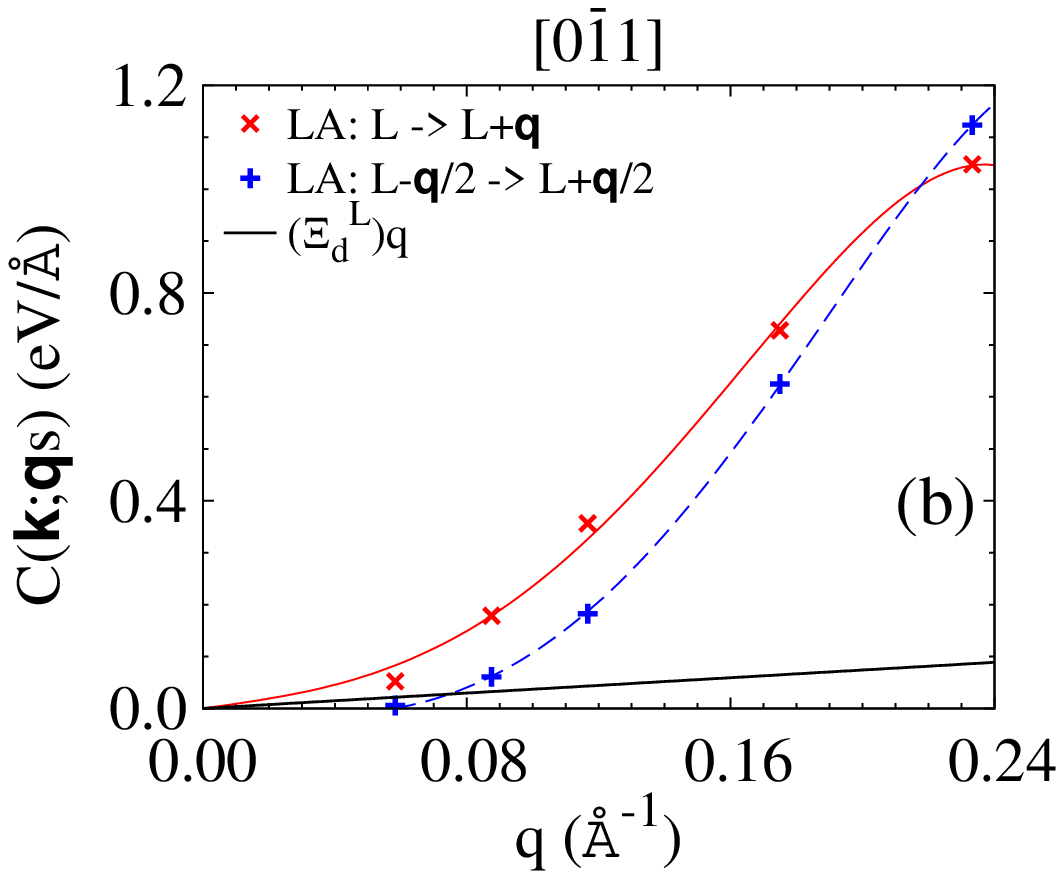} \hspace{3mm}
\includegraphics[width=0.31\linewidth]{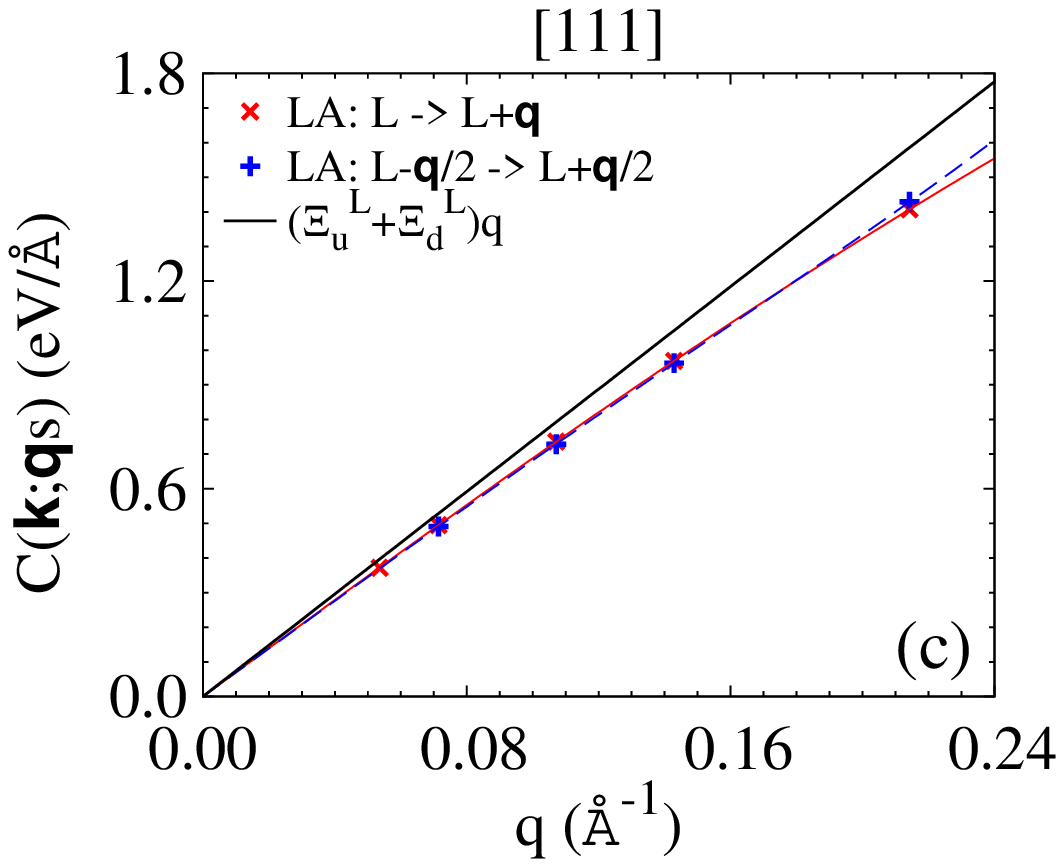}  \\
\includegraphics[width=0.31\linewidth]{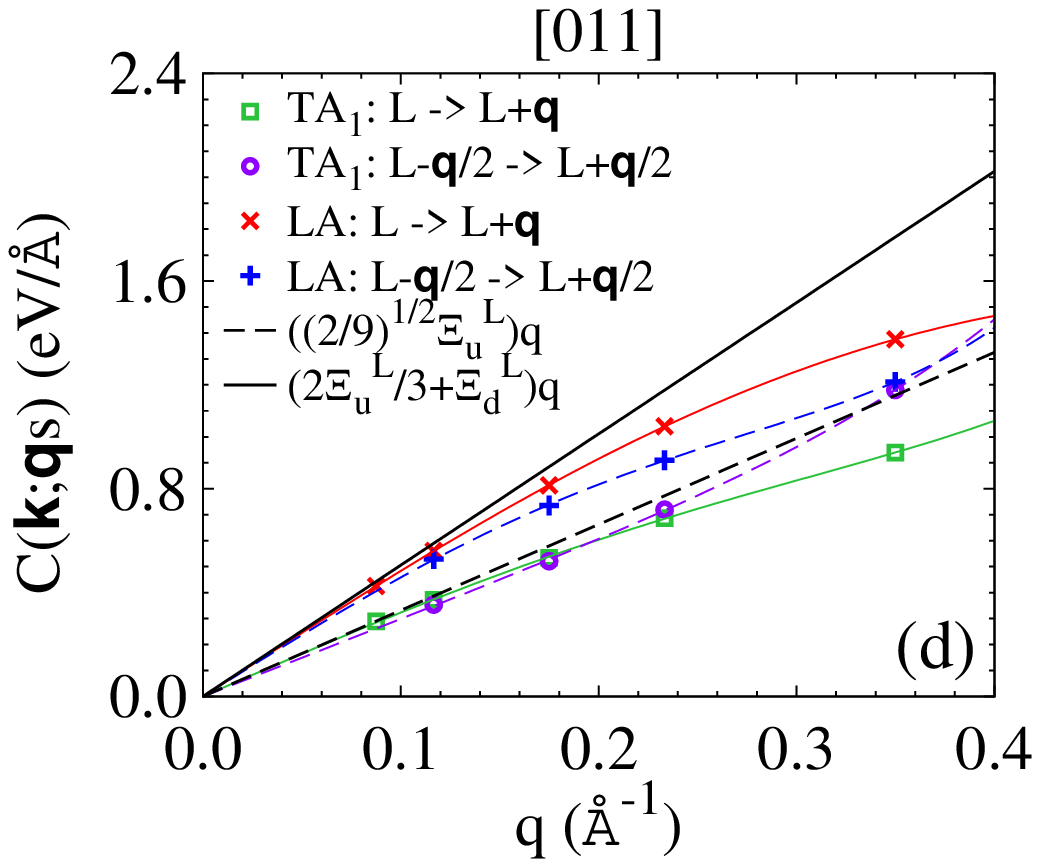} \hspace{10mm}
\includegraphics[width=0.31\linewidth]{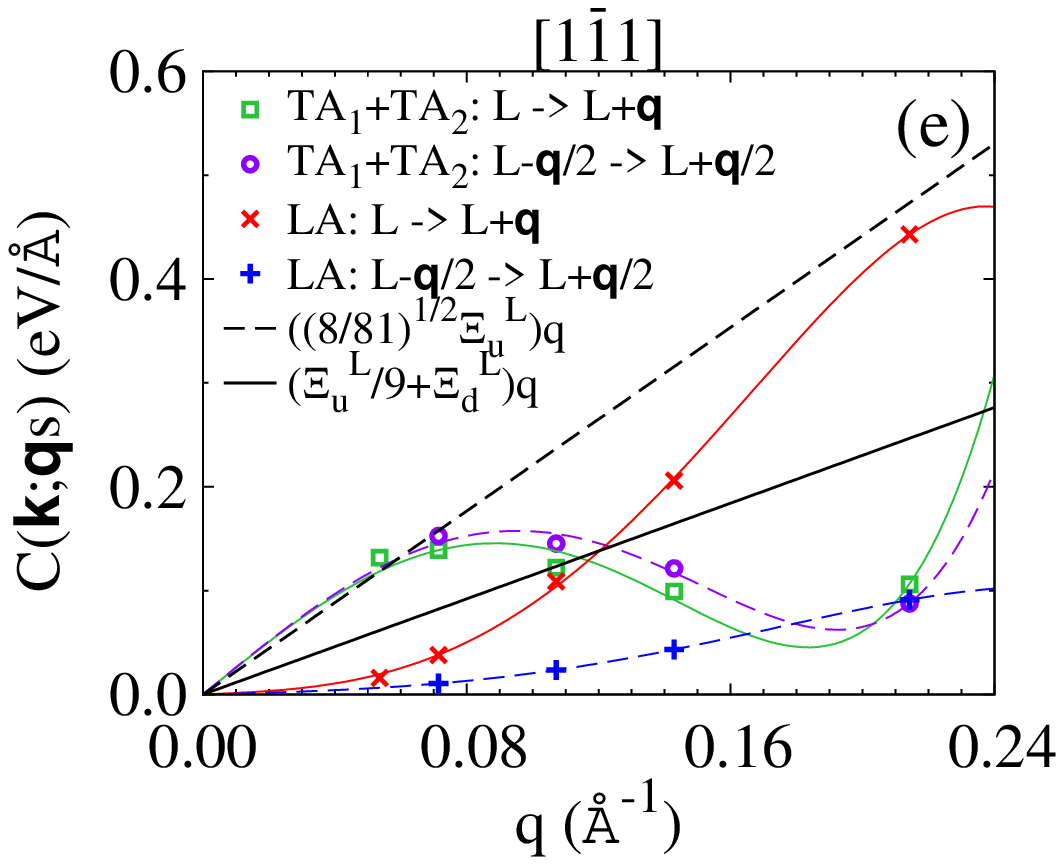} 
\end{center}
\caption{Calculated intravalley electron-phonon matrix elements C(\vect{k}; \vect{q}s) for PbTe as a function of phonon momentum $|\vect{q}|$, for transitions $\text{L}\rightarrow\text{L}+\vect{q}$ via a longitudinal acoustic (LA) phonon (red crosses) or via a transverse acoustic (TA) phonon (green squares), and for transitions $\text{L}-\vect{q}/2\rightarrow\text{L}+\vect{q}/2$ via an LA phonon (blue pluses) or via a TA phonon (purple circles). For $\vect{q}\rightarrow 0$, these yield: (a) $C(\vect{k}; \vect{q}s)\approx \frac{1}{3}\Xi_u^L|\vect{q}|$ for a TA phonon along $[010]$, and $C(\vect{k}; \vect{q}s)\approx (\Xi_d^L + \frac{1}{3}\Xi_u^L)|\vect{q}|$ for an LA phonon along $[010]$, (b) $C(\vect{k}; \vect{q}s)\approx \Xi_{d}^{\text{L}}|\vect{q}|$ for an LA phonon along $[0\bar{1}1]$, (c) $C(\vect{k}; \vect{q}s)\approx (\Xi_d^L + \Xi_u^L)|\vect{q}|$ for an LA phonon along $[111]$, (d): $C(\vect{k}; \vect{q}s)\approx \sqrt{2/9}~\Xi_u^L|\vect{q}|$ for one of the TA phonons along $[011]$, and $C(\vect{k}; \vect{q}s)\approx (\Xi_d^L + \frac{2}{3}\Xi_u^L)|\vect{q}|$ for an LA phonon along $[011]$, and (e) $C(\vect{k}; \vect{q}s)\approx \sqrt{8/81}~\Xi_u^L|\vect{q}|$ for the sum of the two TA phonons along $[1\bar{1}1]$, and $C(\vect{k}; \vect{q}s)\approx (\Xi_d^L + \frac{1}{9}\Xi_u^L)|\vect{q}|$ for an LA phonon along $[1\bar{1}1]$. 5\ts{th} order odd-polynomial fits to the matrix elements are shown with solid lines for $\text{L}\rightarrow\text{L}+\vect{q}$ and dashed lines for $\text{L}-\vect{q}/2\rightarrow\text{L}+\vect{q}/2$. Black solid and dashed lines correspond to the linear fits using the extracted values of $\Xi_u^{\text{L}} = 7.0$ eV and $\Xi_d^{\text{L}} = 0.4$ eV for LA and TA phonons, respectively.}
\label{ch5_xiall}
\end{figure*} 

We calculate the desired electron-phonon matrix elements due to acoustic vibrations by subtracting the Fr\"ohlich contribution from the matrix elements $H_{nn}(\vect{k}; \vect{q}s)$ computed using DFPT:
\begin{equation}
C(\vect{k}; \vect{q}s) = \frac{|H_{nn}(\vect{k}; \vect{q}s)| - |H^{F}_{nn}(\vect{k}; \vect{q}s)|}{\bra{u_{n\vect{k}+\vect{q}}}\ket{u_{n\vect{k}}}_{\text{uc}}}.
\label{ch5_elph_final}
\end{equation}
\noindent In the long wavelength approximation, both total and Fr\"ohlich matrix elements are imaginary if the electronic wavefunctions and phonon eigenvectors are real~\cite{Ridley1999,Ridley1999a}. However, we found that the total matrix elements acquire an arbitrary phase in the DFPT calculations. To avoid any phase mismatch between the DFPT and Fr\"ohlich matrix elements that we calculate aposteriori, we subtract their absolute values, which is a reasonable approximation when $\vect{q}\rightarrow 0$. We found that the values of deformation potentials of PbTe due to the Fr\"ohlich interaction for the longitudinal acoustic phonons range between $0$ to $0.3$ eV along the five high symmetry $\vect{q}$ directions. To be consistent with the standard definition of deformation potentials~\cite{Ridley1999}, we also remove the wavefunction overlap integral from the resulting electron-phonon matrix element, see Eq.~\eqref{ch5_elph_final}. The impact of the wavefunction overlap on our calculated deformation potentials is small since its values become close to 1 as $\vect{q}\rightarrow 0$. 

We plot the values of $C(\vect{k}; \vect{q}s)$ for PbTe as a function of $|\vect{q}|$ for transverse and longitudinal acoustic modes along high symmetry directions in Fig.~\ref{ch5_xiall}, including both $\text{L}\rightarrow\text{L}+\vect{q}$ and $\text{L}-\vect{q}/2\rightarrow\text{L}+\vect{q}/2$ transitions. Fig.~\ref{ch5_xiall} shows that DFPT calculations do not perfectly capture the linear regime as ${\bf q}\rightarrow 0$ for the relatively large values of $|{\bf q}|$ we used. As the $\vect{q}$ vector becomes larger, the deviations of $C(\vect{k}; \vect{q}s)$ from linearity, and thus from the deformation potential approximation, become larger. Similarly, the difference between the $\text{L}\rightarrow\text{L}+\vect{q}$ and $\text{L}-\vect{q}/2\rightarrow\text{L}+\vect{q}/2$ matrix elements for the same direction and polarization increases with $\vect{q}$.

\subsection{\textit{Determining deformation potentials}}

We calculate the values of $H_{ep}/q$ numerically by taking the linear term of the 5\ts{th} order odd-polynomial fit to $C(\vect{k}; \vect{q}s)$ versus $\vect{q}$ for each direction and phonon polarization of interest, given in Table~\ref{table5_directions}. We use an odd-polynomial fit because the Taylor expansion of $H_{nn}(\vect{k}; \vect{q}s)$ for PbTe contains only the odd terms of $\vect{q}$. This is due to the fact that PbTe has an inversion symmetry with respect to each atom, which reverses the sign of its intravalley acoustic phonon eigenvectors \cite{Dresselhaus2008}, while leaving the electron-phonon matrix elements invariant.

\bgroup
\def\arraystretch{1.25}
\begin{table}[htb!]
\begin{center}
\begin{threeparttable}
\begin{tabularx}{0.49\textwidth}{@{}l@{}lC@{}@{}r@{}r}
\hline \hline
$\vect{q}$                               & mode      & Deformation                           & ~~~~~L$\rightarrow$L+$\vect{q}$ & ~~~~~L$-\frac{\vect{q}}{2}\rightarrow$L+$\frac{\vect{q}}{2}$ \\
                                                &                              &                                                                                          & (eV)             & (eV)                \\  [1ex] \hline 
$[010]$                                   & TA                         & $\frac{\Xi_{u}^{\text{L}}}{3}$                                       & $2.27$         & $2.29$           \\ [1ex]
                                                & LA                         & $\Xi_{d}^{\text{L}}+\frac{\Xi_{u}^{\text{L}}}{3}$      & $4.78$         & $4.50$           \\ [1ex] \hline
$[0\bar{1}1]$ \hspace{3mm} & LA                         & $\Xi_{d}^{\text{L}}$                                                       & $0.88$         & $-0.66$         \\ [1ex] \hline
$[111]$                                   & LA                         & $\Xi_{d}^{\text{L}}+\Xi_{u}^{\text{L}}$                      & $6.98$         & $6.95$           \\ [1ex] \hline
$[110]$                                   & TA$_1$                 &  $\sqrt{\frac{2}{9}}\Xi_{u}^{\text{L}}$                         & $3.31$         & $2.98$           \\ [1ex]
                                                & LA                         & $\Xi_{d}^{\text{L}}+\frac{2\Xi_{u}^{\text{L}}}{3}$    & $4.92$         & $4.78$           \\ [1ex] \hline
$[1\bar{1}1]$                         & TA$_1$+TA$_2$  & $\sqrt{\frac{8}{81}}\Xi_{u}^{\text{L}}$                        & $2.58$         & $2.61$           \\ [1ex]
                                                & LA                         & $\Xi_{d}^{\text{L}}+\frac{\Xi_{u}^{\text{L}}}{9}$      & $0.13$        & $0.10$          \\ [1ex] \hline                
\hline
\end{tabularx}
\end{threeparttable}
\end{center}
\caption{Linear terms of the 5\ts{th} order odd-polynomial fits to the electron-phonon matrix elements as defined in Eq.~\eqref{ch5_elph_final} versus $\vect{q}$ for acoustic phonons along the five high-symmetry directions of $\vect{q}$ given in Table~\ref{table5_directions}, for both the
$\text{L}\rightarrow\text{L}+\vect{q}$ and $\text{L}-\vect{q}/2\rightarrow\text{L}+\vect{q}/2$
transitions.}
\label{table5_slopes}
\end{table}
\egroup

We first calculate $\Xi_u^{\text{L}}$ by performing a linear regression fit to the linear terms extracted by fitting only to transverse acoustic matrix elements along the five symmetry directions, see Table~\ref{table5_slopes}. We apply equal weighting to all directions of $\vect{q}$ with a non-zero linear term, with the $\text{L}\rightarrow\text{L}+\vect{q}$ and $\text{L}-\vect{q}/2\rightarrow\text{L}+\vect{q}/2$ matrix elements also weighted equally. This procedure yields the value of $\Xi_u^{\text{L}}$ = 7.0 eV.

We next calculate $\Xi_d^{\text{L}}$ by performing a linear regression fit to the linear terms extracted by fitting only to longitudinal acoustic matrix elements along the five symmetry directions. Enforcing $\Xi_u^{\text{L}}$ = 7.0 eV for all directions, this procedure yields a value of $\Xi_d^{\text{L}}$ = 0.4 eV. The linear fits using these values of $\Xi_u^{\text{L}}$ and $\Xi_d^{\text{L}}$ to the electron-phonon matrix elements for all considered directions and phonon polarizations are shown in Fig.~\ref{ch5_xiall}. Setting $\Xi_u^{\text{L}}$ = 7.0 eV and fitting $\Xi_d^{\text{L}}$ for each $\vect{q}$ direction independently yields values ranging from $-0.7$ to $2.3$ eV. From the mean square displacement of these values, we estimate that the error in our calculated $\Xi_d^{\text{L}}$ is $\sim \pm 1$ eV.

\section{Appendix B: Deformation potentials from band shifts}

We also compute the acoustic deformation potentials of PbTe using the approach of Van de Walle and Martin \cite{VandeWalle1986, VandeWalle1989, VandeWalle1987, VandeWalle1989a}. The calculation of $\Xi_u^{\text{L}}$ is relatively straightforward, since we only need to find the energy splitting between degenerate valleys of bulk due to strain. On the other hand, computing $\Xi_d^{\text{L}}$ is significantly more involved as it requires knowledge of the conduction band energy shifts under strain on an absolute scale \cite{Verges1982}. We note that this method relies on calculating the band shifts under strain using the finite difference method in finite size supercells. These shifts correspond to acoustic deformation potentials only in the limit of infinitely large supercells.

Under uniaxial strain along the $[111]$ direction, the L valleys of a cubic material become non-degenerate and shift in energy proportional to the uniaxial deformation potential and the magnitude of the applied strain. These shifts in energy at the L point may be derived from Eq.~\eqref{ch5_elphHamiltonian_final}, which yields the change in energy of the conduction bands \cite{VandeWalle1986, VandeWalle1989} as:
\begin{equation}
\Delta E_c^{[111]} = 2\Xi_{u}^{\text{L}}\varepsilon_{xy},
\label{ch5_xiu_a}
\end{equation}
\begin{equation}
\Delta E_c^{[\bar{1}11], [1\bar{1}1], [11\bar{1}]} =  -\frac{2}{3}\Xi_{u}^{\text{L}}\varepsilon_{xy}.
\label{ch5_xiu_b}
\end{equation}

\noindent The uniaxial deformation potential then reads:
\begin{equation}
\Xi_{u}^{\text{L}} = \frac{3 \Delta E_{\text{L}}}{8 \varepsilon_{xy}},
\label{xi_u_strain}
\end{equation}

\noindent where $\Delta E_{\text{L}}$ is the energy difference between the $[111]$ and the other L valleys of the strained material. 

To calculate this energy splitting from DFT, we strain the lattice constant parallel to the trigonal $[111]$ axis, $a_{0, \parallel}$, and hold it constant at a value $a_{\parallel}=1.0002\times a_{0, \parallel}$. We then contract the lattice constant perpendicular to the $[111]$ axis, $a_{0,\perp}$, to a value $a_{\perp}$ such that the volume of the primitive cell remains constant. Strain is then defined as \cite{VandeWalle1986, VandeWalle1989}:
\begin{equation}
\varepsilon_{xy} =-\frac{1}{3} \left( \frac{a_{\perp}-a_{0,\perp}}{a_{0,\perp}}-\frac{a_{\parallel}-a_{0,\parallel}}{a_{0,\parallel}} \right).
\end{equation}
We calculate the energy difference between the non-degenerate L valleys under this strain using DFT, and then compute $\Xi_{u}^{\text{L}}$ using Eq.~\eqref{xi_u_strain}. We note that we needed to use small values of strain, $\sim 0.02\--0.1$\%, to obtain converged $\Xi_{u}^{\text{L}}$ values.

The dilatation deformation potential of PbTe can be determined from the shift of the conduction band minimum at L for a dilatation $\varepsilon$ along the $[001]$ direction, which can be found using Eq.~\eqref{ch5_elphHamiltonian_final} \cite{VandeWalle1986, VandeWalle1989}:
\begin{equation}
\Delta E_c^{\text{L}} = \left( \Xi_d^{\text{L}} + \frac{1}{3}\Xi_u^{\text{L}} \right) \varepsilon = a_c \varepsilon.
\label{ac}
\end{equation}
\noindent Here $a_c$ is the hydrostatic deformation potential of the conduction band, and refers to changes of the conduction band states under strain on an absolute scale \cite{Verges1982}. However, the zero of energy is undefined for a bulk crystal due to the long range nature of the Coulomb interaction, resulting in no intrinsic reference for electronic states \cite{Kleinman1981}. To resolve this issue, we calculate $a_c$ following the procedure of Van de Walle and Martin \cite{VandeWalle1986, VandeWalle1989}.

We start by calculating the conduction band energies at the L point for bulk unstrained PbTe and PbTe strained along the $[001]$ direction, $E_c^{\text{L}}(0)$ and $E_c^{\text{L}}(\varepsilon)$ respectively. These DFT calculations also yield the total potential acting on electrons, defined as the sum of ionic, Hartree, and exchange-correlation potentials \cite{VandeWalle1986}, the averages of which we denote as $V(0)$ and $V(\varepsilon)$. Next, we construct a heterostructure which consists of unstrained and strained materials meeting at a $(001)$ interface, which allows us to obtain the average total potential energies of either region, $V'(0)$ and $V'(\varepsilon)$ respectively, on an absolute scale. We then align the total potentials of the bulk calculations to those of the heterostructure, yielding the absolute splitting of the conduction band energies. The hydrostatic deformation potential for the conduction band can then be computed using:
\begin{equation}
\begin{aligned}
a_c =  \frac{\Delta E_c^{\text{L}}}{\varepsilon} =  \frac{1}{\varepsilon} \bigg( & \big( E_c^{\text{L}}(\varepsilon) - V(\varepsilon) \big) - \big(E_c^{\text{L}}(0) - V(0) \big) \\ & + \big( V'(\varepsilon)  - V'(0) \big)\bigg). 
\end{aligned}
\label{ch5_align}
\end{equation}
We calculate the value of $\Xi_d^{\text{L}}$ using Eqs.~\eqref{ac} and \eqref{ch5_align}, using the value of $\Xi_u^{\text{L}}$ obtained from the strain-induced splitting of the L valley. 

 \begin{figure}[!htp]
\begin{center}
\includegraphics[width=0.85\linewidth]{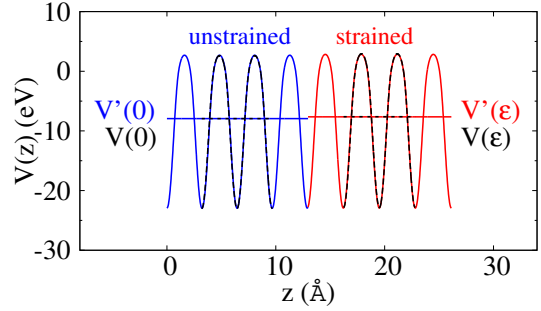}
\end{center}
\caption{Total average potential $V(z)$ for a 16-atom PbTe heterostructure consisting of unstrained and strained regions meeting at a (001) interface, calculated using the HSE03 hybrid functional including spin orbit interaction. The solid blue (red) lines show the total average potential of the unstrained (strained) region, with average potential energy $V'(0)$ ($V'(\varepsilon)$). The total average potential for each value of $z$ in the bulk unstrained and strained material are shown with dashed black lines, aligned to that of the heterostructure so that their average potential energies $V(0)$ and $V(\varepsilon)$ satisfy $V(0)=V'(0)$ and $V(\varepsilon)=V'(\varepsilon)$.}
\label{ch5_strainedcartoon}
\end{figure} 

 To calculate the dilatation deformation potential values of PbTe, we construct a $(001)$ interface between the unstrained and strained regions, with $1$\% of tensile strain applied perpendicular to the interface. Our calculations of the bulk unstrained and strained PbTe are performed on the 4-atom supercells whose lattice vectors are $a[1/2,1/2,0]$, $a[1/2,-1/2,0]$ and $a[0,0,1]$, where $a$ is the lattice constant. The heterostructure is then grown by repeating the 4-atom supercells of the unstrained and strained material along the $[001]$ direction to the desired length. The average potential at each value of $z$ along the $[001]$ direction is defined as:
\begin{equation}
V(z) = \frac{2}{a^2} \int \int V(\vect{r}) dx dy,
\end{equation}
\noindent where $a^2/2$ is the cross-section area perpendicular to the $[001]$ direction of the described heterostructures. We show this average potential $V(z)$ for a 16-atom PbTe heterostructure in Fig.~\ref{ch5_strainedcartoon}, highlighting the unstrained and strained regions. The average potentials of bulk unstrained and strained PbTe are overlaid on top (dashed black lines), shifted in energy so that $V(0)=V'(0)$ and $V(\varepsilon)=V'(\varepsilon)$. As expected, there is a deviation of the average potential of the heterostructure from the bulk behaviour in the vicinity of the interface. We assume that the atoms occupy the ideal positions of the bulk unstrained and strained materials for the considered $[100]$ interface. However, this will not be the case for all interfaces, with certain directions producing a change in the internal displacement parameter. This is of particular importance for polar interfaces, where the change in atomic position produces a dipole shift \cite{VandeWalle1987, Kunc1981}.

\begin{table}[htb!]
\begin{center}
\begin{tabularx}{0.49\textwidth}{@{}l*4{>{\centering\arraybackslash}X}@{}}
\hline \hline
\multicolumn{1}{c}{} & \multicolumn{4}{c}{$\Xi_{d}^{\text{L}}$ ($\text{eV})$} \\ 
\cline{2-5}
Supercell                & LDA           & LDA            &  PBE            & HSE03          \\ [1ex] 
size (atoms)                        & (exc. SOI)   & (inc. SOI)     &    (inc. SOI)       &   (inc. SOI)      \\ [1ex] \hline
8                                  & $~~3.67$ & $~~2.41$ & $-1.73$       &  $-1.20$      \\ [1ex]
16                                & $-0.66$     & $-2.32$     & $-0.13$   &  $-0.22$      \\ [1ex]
24                                & $~~0.21$ & $-1.91$     & $~~0.01$   &  $~~0.91$  \\ [1ex] 
32                                & $~~0.35$ & $-1.69$     & $-0.12$   &  $~~0.69$  \\ [1ex]  
40                                & $~~0.39$ & $-1.66$     & $-0.20$   &  $~~0.78$  \\ [1ex] 
48                                & $~~0.47$ & $-1.55$     & $-0.22$   &  $-$             \\ [1ex] 
\hline \hline
\end{tabularx}
\end{center}
\caption{Dilatation deformation potentials for PbTe calculated using the LDA functional excluding/including spin orbit interaction (SOI), and the PBE and HSE03 functionals including SOI, shown for multiple supercell sizes.}
\label{table5_xid_con}
\end{table}

The supercell size introduces periodicity into the heterostructure calculation, and its size must be sufficiently large to restore the bulk behaviour for the average potential away from the interface. However, we found that our calculated values of $a_c$ did not converge smoothly with increasing supercell size. This numerical instability was due to the relatively sparse set of $z$ points defining the average potential $V(z)$, which is determined by the number of Fourier components in the plane wave expansion that is fixed by the energy cutoff ($45$ Ha for the LDA-HGH calculations, and $18.4$ Ha for the PBE/HSE03 PAW calculations). For example, the calculated $V(z)$ values did not necessarily align to the peaks and troughs of $V(z)$. Thus, we performed a cubic spline interpolation of the average potential to generate a sufficiently dense set of $V(z)$ values. We extracted the average potentials $V'(0)$ and $V'(\varepsilon)$ from the center of the unstrained and strained regions of the heterostructure, where the average potential is most bulk-like. In doing so, we defined the region over which the average potential energy is calculated as the central 4-atom subsection of the unstrained and strained parts of the heterostructure. This procedure yielded a better convergence with respect to increasing supercell size, see Table~\ref{table5_xid_con}. From the differences of the dilatation potential values for two largest supecell sizes, we estimate that their error in these calculations is $\sim \pm 0.1$ eV.

%%%%%%%%%%%%%%%%%%%
% Bibliography
\bibliography{thesis}

\end{document}